\pdfoutput=1

\documentclass[12pt,a4paper]{article}

\usepackage{ifthen}
\newboolean{pdflatex}
\setboolean{pdflatex}{true}

\newboolean{articletitles}
\setboolean{articletitles}{true}

\newboolean{uprightparticles}
\setboolean{uprightparticles}{false}

\newboolean{inbibliography}
\setboolean{inbibliography}{false}

\def\paperauthors{LHCb collaboration}
\def\paperasciititle{Search for beautiful tetraquarks in the Ymumu invariant-mass spectrum}
\def\papertitle{Search for beautiful tetraquarks in the $\OneS\mu^+\mu^-$ invariant-mass spectrum}
\def\paperkeywords{{High Energy Physics}, {LHCb}}
\def\papercopyright{\the\year\ CERN for the benefit of the LHCb collaboration}
\def\paperlicence{CC-BY-4.0}
\def\paperlicenceurl{https://creativecommons.org/licenses/by/4.0/}

\usepackage[top=1in, bottom=1.25in, left=1in, right=1in]{geometry}

\usepackage{microtype}
\usepackage{lineno}
\usepackage{xspace}
\usepackage{caption}

\usepackage{graphicx}
\usepackage{color}
\usepackage{colortbl}

\usepackage{amsmath}
\usepackage{amssymb}
\usepackage{amsfonts}
\usepackage{upgreek}

\newcommand*\patchAmsMathEnvironmentForLineno[1]{
\expandafter\let\csname old#1\expandafter\endcsname\csname #1\endcsname
\expandafter\let\csname oldend#1\expandafter\endcsname\csname
end#1\endcsname
 \renewenvironment{#1}
   {\linenomath\csname old#1\endcsname}
   {\csname oldend#1\endcsname\endlinenomath}
}
\newcommand*\patchBothAmsMathEnvironmentsForLineno[1]{
  \patchAmsMathEnvironmentForLineno{#1}
  \patchAmsMathEnvironmentForLineno{#1*}
}
\AtBeginDocument{
\patchBothAmsMathEnvironmentsForLineno{equation}
\patchBothAmsMathEnvironmentsForLineno{align}
\patchBothAmsMathEnvironmentsForLineno{flalign}
\patchBothAmsMathEnvironmentsForLineno{alignat}
\patchBothAmsMathEnvironmentsForLineno{gather}
\patchBothAmsMathEnvironmentsForLineno{multline}
\patchBothAmsMathEnvironmentsForLineno{eqnarray}
}

\usepackage{hyperxmp}

\usepackage[pdftex,
            pdfauthor={\paperauthors},
            pdftitle={\paperasciititle},
            pdfkeywords={\paperkeywords},
            pdfcopyright={Copyright (C) \papercopyright},
            pdflicenseurl={\paperlicenceurl}]{hyperref}

\usepackage[all]{hypcap}

\usepackage{xspace}
\usepackage{upgreek}

\def\lhcb {\mbox{LHCb}\xspace}

\def\MagUp {\mbox{\em Mag\kern -0.05em Up}\xspace}

\ifthenelse{\boolean{uprightparticles}}
{

 \def\Pmu         {\ensuremath{\upmu}\xspace}

 \def\Ppsi        {\ensuremath{\uppsi}\xspace}

 \def\PDelta      {\ensuremath{\Delta}\xspace}
 \def\PXi      {\ensuremath{\Xi}\xspace}
 \def\PLambda      {\ensuremath{\Lambda}\xspace}
 \def\PSigma      {\ensuremath{\Sigma}\xspace}
 \def\POmega      {\ensuremath{\Omega}\xspace}
 \def\PUpsilon      {\ensuremath{\Upsilon}\xspace}

 \def\PB      {\ensuremath{\mathrm{B}}\xspace}
 
 \def\PD      {\ensuremath{\mathrm{D}}\xspace}

 \def\PJ      {\ensuremath{\mathrm{J}}\xspace}
 \def\PK      {\ensuremath{\mathrm{K}}\xspace}

 \def\Pb      {\ensuremath{\mathrm{b}}\xspace}
 \def\Pc      {\ensuremath{\mathrm{c}}\xspace}

 \def\Pi      {\ensuremath{\mathrm{i}}\xspace}

 \def\Pp      {\ensuremath{\mathrm{p}}\xspace}

 \def\Ps      {\ensuremath{\mathrm{s}}\xspace}

}
{

 \def\Pmu         {\ensuremath{\mu}\xspace}

 \def\Ppsi        {\ensuremath{\psi}\xspace}
 
 \mathchardef\PDelta="7101
 \mathchardef\PXi="7104
 \mathchardef\PLambda="7103
 \mathchardef\PSigma="7106
 \mathchardef\POmega="710A
 \mathchardef\PUpsilon="7107
 
 \def\PB      {\ensuremath{B}\xspace}
 
 \def\PD      {\ensuremath{D}\xspace}

 \def\PJ      {\ensuremath{J}\xspace}
 \def\PK      {\ensuremath{K}\xspace}

 \def\Pb      {\ensuremath{b}\xspace}
 \def\Pc      {\ensuremath{c}\xspace}

 \def\Pi      {\ensuremath{i}\xspace}

 \def\Pp      {\ensuremath{p}\xspace}

 \def\Ps      {\ensuremath{s}\xspace}

}

\makeatletter
\ifcase \@ptsize \relax
  \newcommand{\miniscule}{\@setfontsize\miniscule{4}{5}}
\or
  \newcommand{\miniscule}{\@setfontsize\miniscule{5}{6}}
\or
  \newcommand{\miniscule}{\@setfontsize\miniscule{5}{6}}
\fi
\makeatother

\DeclareRobustCommand{\optbar}[1]{\shortstack{{\miniscule (\rule[.5ex]{1.25em}{.18mm})}
  \\ [-.7ex] $#1$}}

\def\mup        {{\ensuremath{\Pmu^+}}\xspace}
\def\mun        {{\ensuremath{\Pmu^-}}\xspace}
\def\mumu       {{\ensuremath{\Pmu^+\Pmu^-}}\xspace}

\def\squark    {{\ensuremath{\Ps}}\xspace}

\def\cquark    {{\ensuremath{\Pc}}\xspace}

\def\bquark    {{\ensuremath{\Pb}}\xspace}

  \def\Kbar    {{\kern 0.2em\overline{\kern -0.2em \PK}{}}\xspace}

\def\KorKbar    {\kern 0.18em\optbar{\kern -0.18em K}{}\xspace}

  \def\Dbar    {{\kern 0.2em\overline{\kern -0.2em \PD}{}}\xspace}

\def\DorDbar    {\kern 0.18em\optbar{\kern -0.18em D}{}\xspace}

\def\B       {{\ensuremath{\PB}}\xspace}
\def\Bbar    {{\ensuremath{\kern 0.18em\overline{\kern -0.18em \PB}{}}}\xspace}

\def\BorBbar    {\kern 0.18em\optbar{\kern -0.18em B}{}\xspace}

\def\Bd      {{\ensuremath{\B^0}}\xspace}
\def\Bs      {{\ensuremath{\B^0_\squark}}\xspace}

\def\jpsi     {{\ensuremath{{\PJ\mskip -3mu/\mskip -2mu\Ppsi\mskip 2mu}}}\xspace}

  \def\Y#1S{\ensuremath{\PUpsilon{(#1S)}}\xspace}
\def\OneS  {{\Y1S}}
\def\TwoS  {{\Y2S}}
\def\ThreeS{{\Y3S}}
\def\FourS {{\Y4S}}

\def\proton      {{\ensuremath{\Pp}}\xspace}

\def\Lbar        {{\ensuremath{\kern 0.1em\overline{\kern -0.1em\PLambda}}}\xspace}
\def\LorLbar    {\kern 0.18em\optbar{\kern -0.18em \PLambda}{}\xspace}

\newcommand{\decay}[2]{\ensuremath{#1\!\to #2}\xspace}

\def\to                 {\ensuremath{\rightarrow}\xspace}

\def\AT#1     {\ensuremath{A_{\mathrm{T}}^{#1}}\xspace}

\def\C#1      {\ensuremath{\mathcal{C}_{#1}}\xspace}
\def\Cp#1     {\ensuremath{\mathcal{C}_{#1}^{'}}\xspace}
\def\Ceff#1   {\ensuremath{\mathcal{C}_{#1}^{\mathrm{(eff)}}}\xspace}
\def\Cpeff#1  {\ensuremath{\mathcal{C}_{#1}^{'\mathrm{(eff)}}}\xspace}
\def\Ope#1    {\ensuremath{\mathcal{O}_{#1}}\xspace}
\def\Opep#1   {\ensuremath{\mathcal{O}_{#1}^{'}}\xspace}

\newcommand{\tev}{\ifthenelse{\boolean{inbibliography}}{\ensuremath{~T\kern -0.05em eV}}{\ensuremath{\mathrm{\,Te\kern -0.1em V}}}\xspace}
\newcommand{\tevfix}{\ensuremath{\mathrm{\,Te\kern -0.1em V}}\xspace}
\newcommand{\gev}{\ensuremath{\mathrm{\,Ge\kern -0.1em V}}\xspace}
\newcommand{\mev}{\ensuremath{\mathrm{\,Me\kern -0.1em V}}\xspace}
\newcommand{\kev}{\ensuremath{\mathrm{\,ke\kern -0.1em V}}\xspace}
\newcommand{\ev}{\ensuremath{\mathrm{\,e\kern -0.1em V}}\xspace}
\newcommand{\gevc}{\ensuremath{{\mathrm{\,Ge\kern -0.1em V\!/}c}}\xspace}
\newcommand{\mevc}{\ensuremath{{\mathrm{\,Me\kern -0.1em V\!/}c}}\xspace}
\newcommand{\gevcc}{\ensuremath{{\mathrm{\,Ge\kern -0.1em V\!/}c^2}}\xspace}
\newcommand{\gevgevcccc}{\ensuremath{{\mathrm{\,Ge\kern -0.1em V^2\!/}c^4}}\xspace}
\newcommand{\mevcc}{\ensuremath{{\mathrm{\,Me\kern -0.1em V\!/}c^2}}\xspace}

\def\mum  {\ensuremath{{\,\upmu\mathrm{m}}}\xspace}

\def\fb   {\ensuremath{\mbox{\,fb}}\xspace}
\def\invfb   {\ensuremath{\mbox{\,fb}^{-1}}\xspace}

\def\gsim{{~\raise.15em\hbox{$>$}\kern-.85em
          \lower.35em\hbox{$\sim$}~}\xspace}
\def\lsim{{~\raise.15em\hbox{$<$}\kern-.85em
          \lower.35em\hbox{$\sim$}~}\xspace}

\def\sqs   {\ensuremath{\protect\sqrt{s}}\xspace}

\def\ptot       {\mbox{$p$}\xspace}
\def\pt         {\mbox{$p_{\mathrm{ T}}$}\xspace}

\def\evtgen     {\mbox{\textsc{EvtGen}}\xspace}

\def\geant      {\mbox{\textsc{Geant4}}\xspace}

\def\photos     {\mbox{\textsc{Photos}}\xspace}

\def\pythia     {\mbox{\textsc{Pythia}}\xspace}

\def\tell1  {TELL1\xspace}
\def\ukl1   {UKL1\xspace}

\usepackage{cite}
\usepackage{mciteplus}
\usepackage{longtable}

\begin{document}

\renewcommand{\thefootnote}{\fnsymbol{footnote}}
\setcounter{footnote}{1}

\begin{titlepage}
\pagenumbering{roman}

\vspace*{-1.5cm}
\centerline{\large EUROPEAN ORGANIZATION FOR NUCLEAR RESEARCH (CERN)}
\vspace*{1.5cm}
\noindent
\begin{tabular*}{\linewidth}{lc@{\extracolsep{\fill}}r@{\extracolsep{0pt}}}
\vspace*{-1.5cm}\mbox{\!\!\!\includegraphics[width=.14\textwidth]{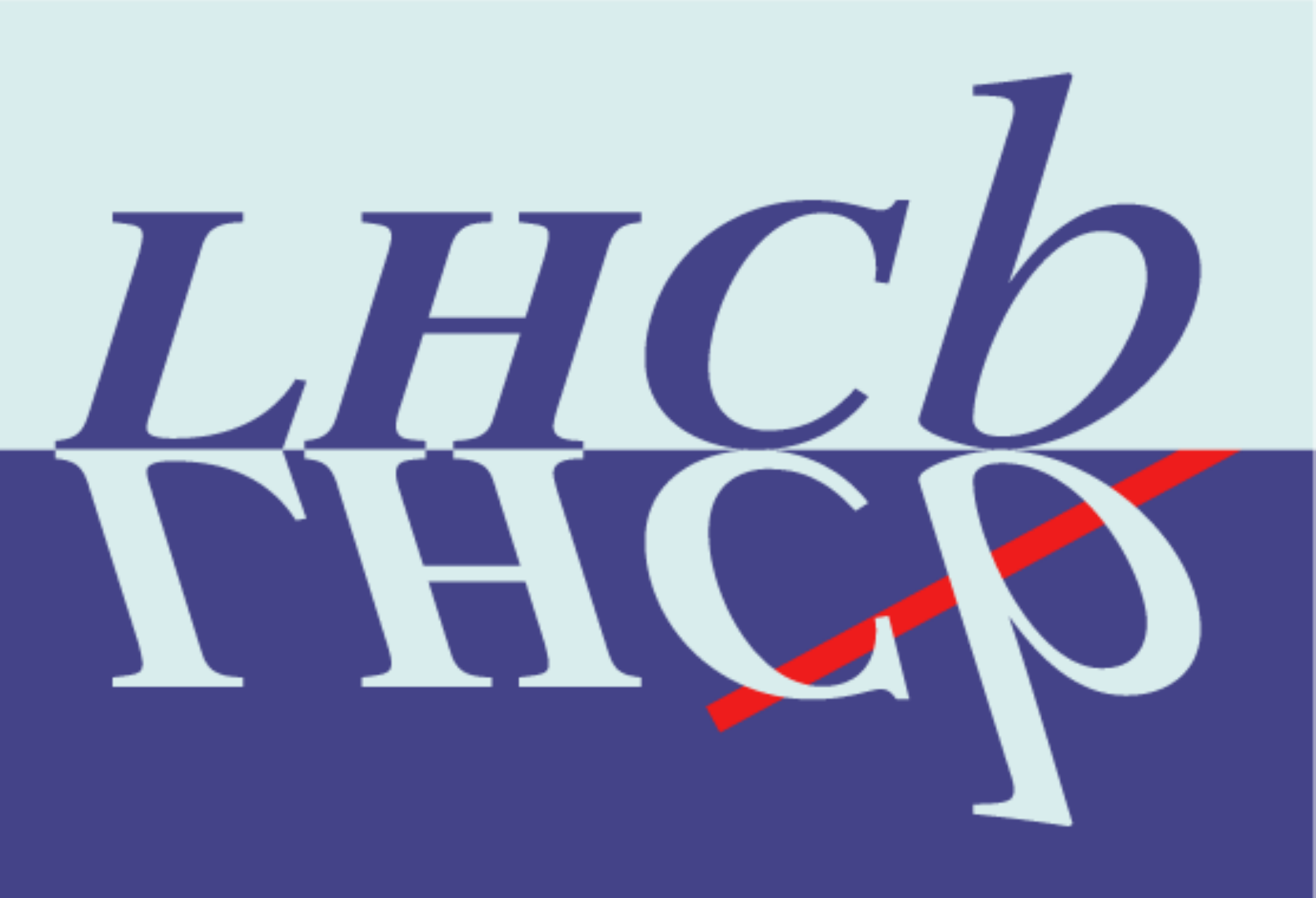}} & &
\\
 & & CERN-EP-2018-163 \\
 & & LHCb-PAPER-2018-027 \\
 & & November 7, 2018 \\
 & & \\
\end{tabular*}

\vspace*{4.0cm}

{\normalfont\bfseries\boldmath\huge
\begin{center}
  \papertitle
\end{center}
}

\vspace*{2.0cm}

\begin{center}
\paperauthors\footnote{Authors are listed at the end of this paper.}
\end{center}

\vspace{\fill}

\begin{abstract}
  \noindent
    The $\OneS\mumu$ invariant-mass distribution is investigated for a possible exotic meson state
    composed of two $b$ quarks and two $\overline{b}$ quarks,
    $X_{b\overline{b}b\overline{b}}$.  The analysis is based on a data sample
    of $pp$ collisions recorded with the LHCb detector at
    centre-of-mass energies $\sqrt{s} = 7$, 8 and 13\tev,
    corresponding to an integrated luminosity of 6.3\invfb.
    No significant excess is found, and
    upper limits are set on the product of the production cross-section and the branching
    fraction as functions of the mass of the
    $X_{b\overline{b}b\overline{b}}$ state.  The limits are set in the fiducial
    volume where all muons have pseudorapidity in the range $[2.0,5.0]$,
    and the $X_{b\overline{b}b\overline{b}}$ state has rapidity in the range $[2.0,4.5]$
    and transverse momentum less than $15 \gevc$.
\end{abstract}

\vspace*{2.0cm}

\begin{center}
  Published in JHEP {\bf 10} (2018) 086
\end{center}

\vspace{\fill}

{\footnotesize
\centerline{\copyright~\papercopyright. \href{\paperlicenceurl}{\paperlicence} licence.}}
\vspace*{2mm}

\end{titlepage}

\newpage
\setcounter{page}{2}
\mbox{~}

\cleardoublepage

\renewcommand{\thefootnote}{\arabic{footnote}}
\setcounter{footnote}{0}

\pagestyle{plain}
\setcounter{page}{1}
\pagenumbering{arabic}

\section{Introduction}
\label{sec:Introduction}

Since the discovery of the $X(3872)$ state~\cite{Choi:2003ue}, over thirty
exotic hadrons have been observed by several experiments (see
Refs.~\cite{Chen:2016qju,Lebed:2016hpi,Esposito:2016noz,Guo:2017jvc,Ali:2017jda,Olsen:2017bmm}
for recent reviews). Most progress has been seen in
the charmonium sector, where tetraquark (pentaquark) candidates
with masses around 4\gevcc have been found decaying to final states containing
charmonia and are believed to have a
minimal quark content of $c\overline{c}q\overline{q}^\prime\,(c\overline{c}qq^\prime q^{\prime\prime})$, where $q$
refers to a light quark $(u,d,s)$. Two tetraquark states have also been seen in
the bottomonium sector, via their decay to $\PUpsilon\pi$
final states~\cite{Belle:2011aa}.

So far, no exotic hadron that is composed of more than two heavy quarks has been observed.
However, there have recently
been several predictions for the mass and width of an exotic state,
$X_{b\overline{b}b\overline{b}}$ (denoted by $X$ in the following), with quark composition
$b\overline{b}b\overline{b}$~\cite{Heller:1985cb,Berezhnoy:2011xn,Wu:2016vtq,Chen:2016jxd,Karliner:2016zzc,Bai:2016int,Wang:2017jtz,Richard:2017vry,Anwar:2017toa,Vega-Morales:2017pmm,Chen:2018cqz}.
These predictions indicate that the
$X$ state would have a mass in the region
[18.4, 18.8]\gevcc, placing
it close to, but typically below, the $\eta_b\eta_b$
threshold of ${18.798\pm0.005 \gevcc}$~\cite{PDG2016},
which implies that it could decay to  $\PUpsilon \ell^+\ell^-$ ($\ell = e, \mu$)
final states.
Further motivation is provided by the recent
observation of $\OneS\OneS$ production by the CMS collaboration~\cite{Khachatryan:2016ydm}.
Possible search strategies for the
$X$ state have been outlined in
Ref.~\cite{Eichten:2017ual}, and
the product of its production cross-section at the LHC and the branching fraction to four muons is estimated to be of
$\mathcal{O}(1\fb)$. However, recent lattice QCD
calculations do not find evidence for such a state in the hadron
spectrum~\cite{Hughes:2017xie}.

The current paper presents the first search for
this state decaying to $\OneS\mumu$ through a study of the four-muon invariant-mass distribution,
$m(2\mup2\mun)$, between 17.5 and 20.0\gevcc. The dataset consists of
$\proton\proton$ collision data
recorded by the LHCb experiment at
centre-of-mass energies of $\sqs=7\tev$, $8\tev$ and $13\tev$
between 2011 and 2017.
The corresponding integrated luminosities are
$1.0\invfb$, $2.0\invfb$ and $3.3\invfb$, respectively.
The $\OneS\to\mumu$ decay is used as a normalisation channel to
calculate the $X$ production cross-section relative
to that of the $\OneS$ meson.

\section{Detector and simulation}
\label{sec:Detector}

The \lhcb detector~\cite{Alves:2008zz,LHCb-DP-2014-002} is a single-arm forward
spectrometer covering the \mbox{pseudorapidity} range $2<\eta <5$,
designed for the study of particles containing \bquark or \cquark
quarks. The detector includes a high-precision tracking system
consisting of a silicon-strip vertex detector surrounding the $pp$
interaction region, a large-area silicon-strip detector located
upstream of a dipole magnet with a bending power of about
$4{\mathrm{\,Tm}}$, and three stations of silicon-strip detectors and straw
drift tubes placed downstream of the magnet.
The tracking system provides a measurement of the momentum, \ptot, of charged particles with
a relative uncertainty that varies from 0.5\% at low momentum to 1.0\% at 200\gevc.
The minimum distance of a track to a primary vertex (PV), the impact parameter (IP),
is measured with a resolution of $(15+29/\pt)\mum$,
where \pt is the component of the momentum transverse to the beam, in\,\gevc.
Different types of charged hadrons are distinguished using information
from two ring-imaging Cherenkov detectors.
Photons, electrons and hadrons are identified by a calorimeter system consisting of
scintillating-pad and preshower detectors, an electromagnetic
calorimeter and a hadronic calorimeter. Muons are identified by a
system composed of alternating layers of iron and multiwire
proportional chambers~\cite{LHCb-DP-2012-002}.

Simulated datasets are used to evaluate reconstruction and selection efficiencies of the $\OneS$ and $X$ decays studied in this paper.
In the simulation, $pp$ collisions are generated using
\pythia~\cite{Sjostrand:2007gs,Sjostrand:2006za}
 with a specific \lhcb
configuration~\cite{LHCb-PROC-2010-056}.  Decays of hadronic particles
are described by \evtgen~\cite{Lange:2001uf}, in which final-state
radiation is generated using \photos~\cite{Golonka:2005pn}. The
interaction of the generated particles with the detector, and its response,
are implemented using the \geant
toolkit~\cite{Allison:2006ve, *Agostinelli:2002hh} as described in
Ref.~\cite{LHCb-PROC-2011-006}.
The $X$ state is produced using the same production model as the $\FourS$ meson,
with the mass changed to one of three values in the range $18\,450-18\,830\mevcc$.
The natural width of the $X$ state is assumed to be $1.2\mevcc$ and its decay to the
$\OneS\mup\mun$ final state is modelled by a phase-space distribution.
The kinematic distribution of simulated $X$ particles is shown in the \hyperref[sec:appendix]{Appendix}.

\section{Event selection}
\label{sec:selection}

For both signal and normalisation channels,
the $\decay{\OneS}{\mumu}$ candidates are first required to pass the
trigger~\cite{LHCb-DP-2012-004},
which consists of a hardware stage, based on information from the calorimeter and muon
systems, followed by a software stage, which applies a full event
reconstruction.
At the hardware level, a minimum requirement is placed on the product of the transverse momenta of the two muons. At the software level, requirements are made on
the total and transverse momentum of these muons, the dimuon invariant mass and
on the quality of the dimuon vertex fit.
Additionally, requirements are placed on the track quality of the
muons and on particle identification (PID) quantities of the muons.

In the offline selection, all muons are required
to have $\ptot\in[8,500]\gevc$, $\pt$ larger than $1\gevc$
and $\eta\in[2.0,5.0]$.
Stringent requirements are also applied to muon track-quality and PID quantities to reduce backgrounds from particles that are misidentified as muons. For both signal and normalisation channels, all muons are required to be consistent with originating from a common PV.
The $\OneS\to\mup\mun$ candidates are required to have
invariant masses ${m(\mu^+\mu^-)\in[8.5,11.5]\gevcc}$
and a good vertex-fit quality.

For the $\decay{X}{\OneS\mup\mun}$ decay,
the $\OneS$ candidates are combined with an
additional dimuon pair with a good vertex-fit quality.
In addition to the four-muon vertex fit having
good quality,
the $X$ candidates are required to have invariant masses
${m(2\mu^+2\mu^-)\in[16.0,22.0]\gevcc}$,
rapidities in the range $[2.0,4.5]$ and \pt less
than $15 \gevc$. If a same-charge pair
of muons has an invariant mass less than $220\mevcc$ or an opening angle smaller than 0.002 radians,
then the corresponding $X$ candidate is removed.
This requirement eliminates pairs of muon candidates that are wrongly reconstructed from
one single track.
Candidates are also rejected if the combination of either muon from the $\OneS$ decay with the oppositely charged additional muon has an invariant mass consistent with
that of the $\jpsi$ meson, $m(\mu^+\mu^-)\in[3050,3150]\mevcc$.
The signal sample is a subset of the normalisation sample, smaller by a factor of ${\cal  O}(10^4)$.

Multiple $X$ candidates are seen in approximately $10\,\%$ of events that pass the full selection
and have $m(\mumu)$ within $\pm100\mevcc$ of the known $\OneS$ mass~\cite{PDG2016}. These are mostly due to
the same $\OneS$ candidate being combined with different additional dimuons. These candidates are retained
and treated as combinatorial background.
Events with multiple candidates in the normalization $\OneS$ dataset occur at a negligible level.

\section{Invariant-mass fits}

Unbinned extended maximum-likelihood fits
are made to the $m(2\mup2\mun)$ and $m(\mup\mun)$ distributions to determine $X$ and $\OneS$ yields, respectively.
Fits to three datasets collected at $pp$
centre-of-mass energies of $\sqs=7\tev$ in 2011, $8\tev$ in 2012
and $13\tev$ in 2015--2017 are performed. In addition, a fit is made
to a merged dataset that
combines all $7$, $8$ and $13\tev$ subsets.
In each fit, the combinatorial background component is
described by an exponential function with the
slope and normalisation as free parameters.
Signal components are described by Crystal Ball functions~\cite{Skwarnicki:1986xj}
with the tail parameters fixed to values obtained from fits to the simulated samples.

In fits to the $m(\mumu)$ distributions, contributions from the $\OneS$, $\TwoS$ and $\ThreeS$ states are included.
For the $\OneS$ contribution, the mean, $\mu_{\OneS}$, and width, $\sigma_{\OneS}$, of the shape are free parameters.
For the $\PUpsilon(nS)$ contributions ($n = 2,3$) the means are free parameters while each width is
fixed to that of the $\OneS$ component scaled by the ratio of the $\OneS$ and $\PUpsilon(nS)$ masses. 
The number of candidates of each component is free in each fit.

\begin{figure}
	\center
	\includegraphics[width=0.49\textwidth]{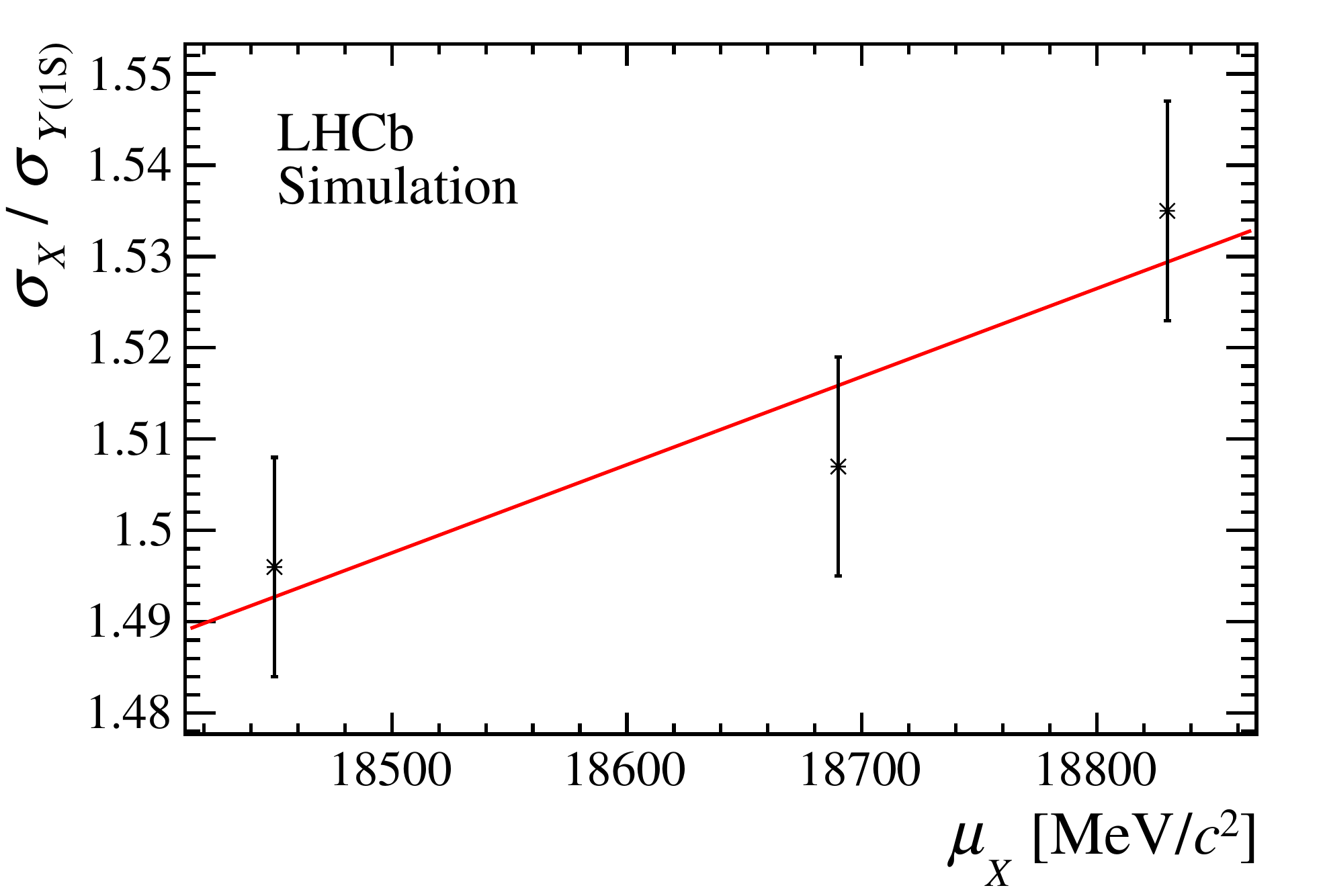}
    \caption{\label{fig:resofit}\small Linear fit to the ratio of the $X$
    and $\OneS$ widths as a function of the $X$ mass as determined from fits to simulated data samples.
    The error bars represent the statistical uncertainty arising from the
    finite size of the simulated samples.
    }
\end{figure}

In the fits to the $m(2\mup2\mun)$ distributions, the mean of the $X$ contribution, $\mu_{X}$,
takes a value in the range $[17.5,20.0]\gevcc$,
while the width, $\sigma_{X}$, is calculated as
the product of the corresponding $\OneS$ resolution and a linear $X$-mass-dependent scaling factor~\cite{LHCb-PAPER-2017-001},
$\sigma_{X}=k(\mu_{X})\times\sigma_{\OneS}$ with $k(\mu_{X}) = p_0 + p_1 (\mu_{X}-18\,690\mevcc)$.
The values of the $\OneS$ resolution and the two coefficients of the linear function are constrained by Gaussian functions.
The constraints on the $\OneS$ resolution, taken from fits to the normalisation datasets, are
$44.00\pm0.05$, $44.307\pm0.035$, {$43.155\pm0.023$} and {$43.766\pm0.018\mevcc$}
for the 7, 8, 13\tev and combined datasets, respectively. The constraints on $p_0$ and $p_1$
are $1.516\pm0.007$ and ${(9.6\pm4.4)\times10^{-5}\,(\mevcc)^{-1}}$, respectively,
evaluated from a fit to the simulated data, as shown in Fig.~\ref{fig:resofit}.
These constraints lead to typical $X$ resolutions in the range $\sim[60,70]\mevcc$.

\begin{figure}
	\centering
	\includegraphics[width=0.49\textwidth]{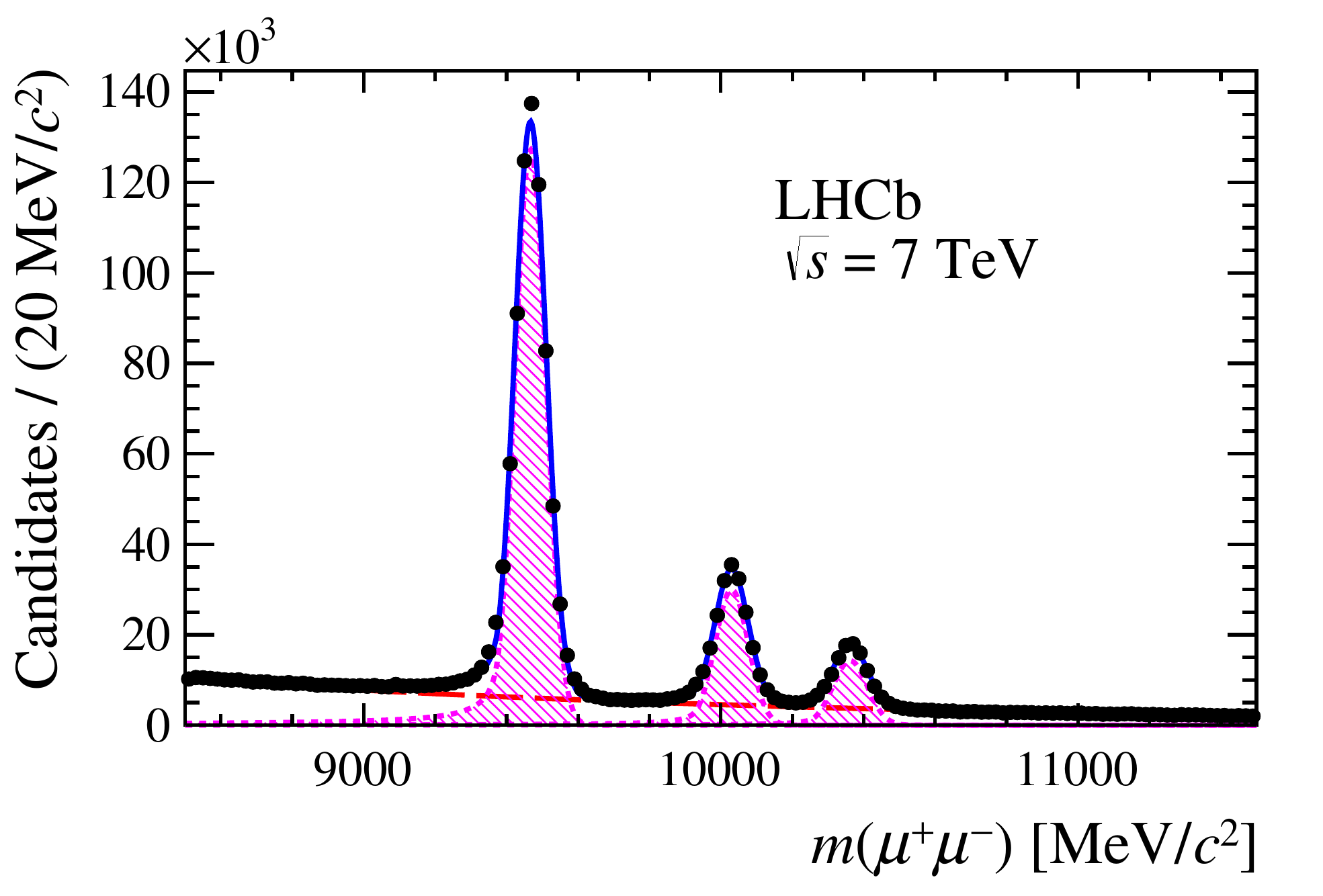}
    \put(-92,85){(a)}
	\includegraphics[width=0.49\textwidth]{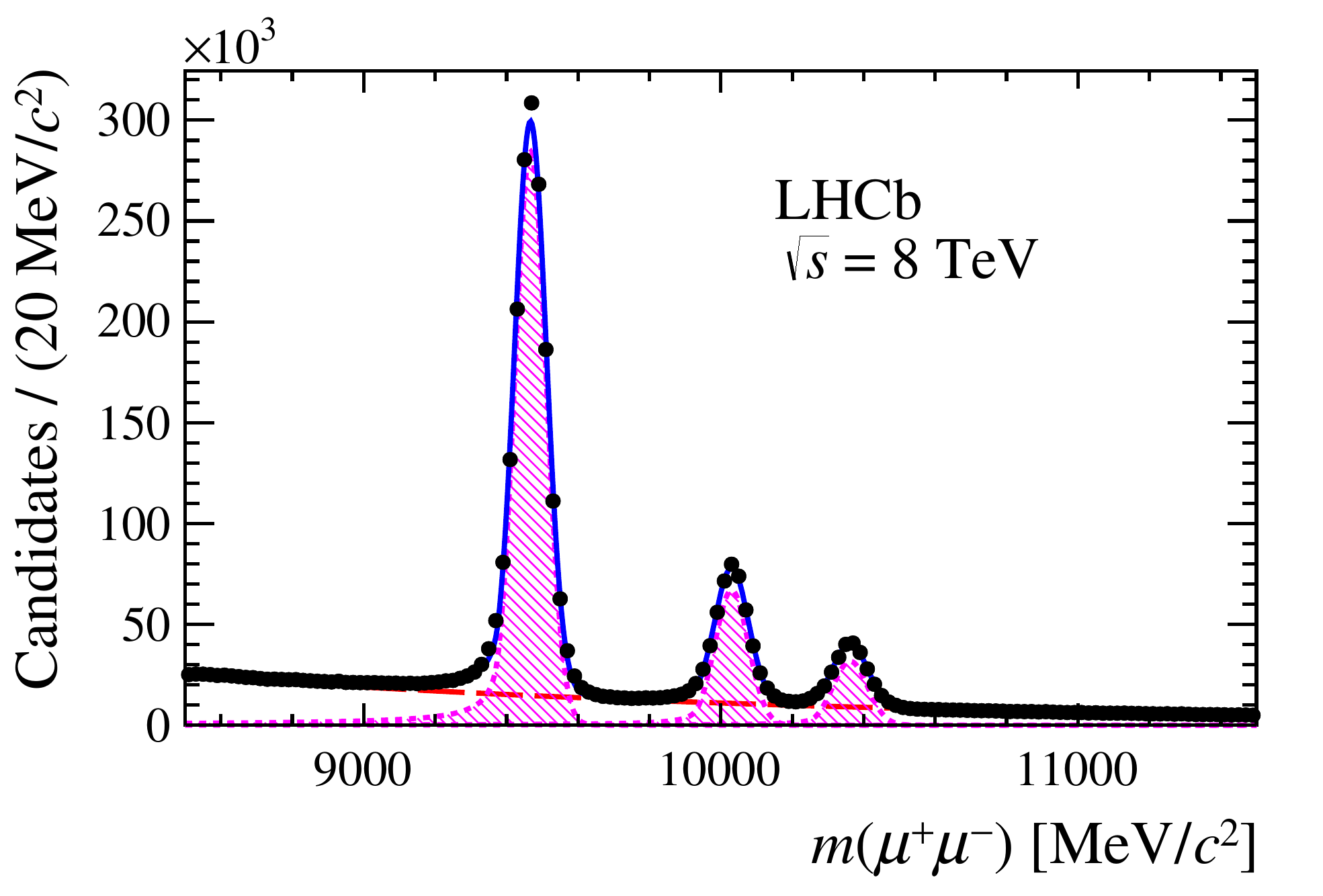}
    \put(-92,85){(b)}\\
	\includegraphics[width=0.49\textwidth]{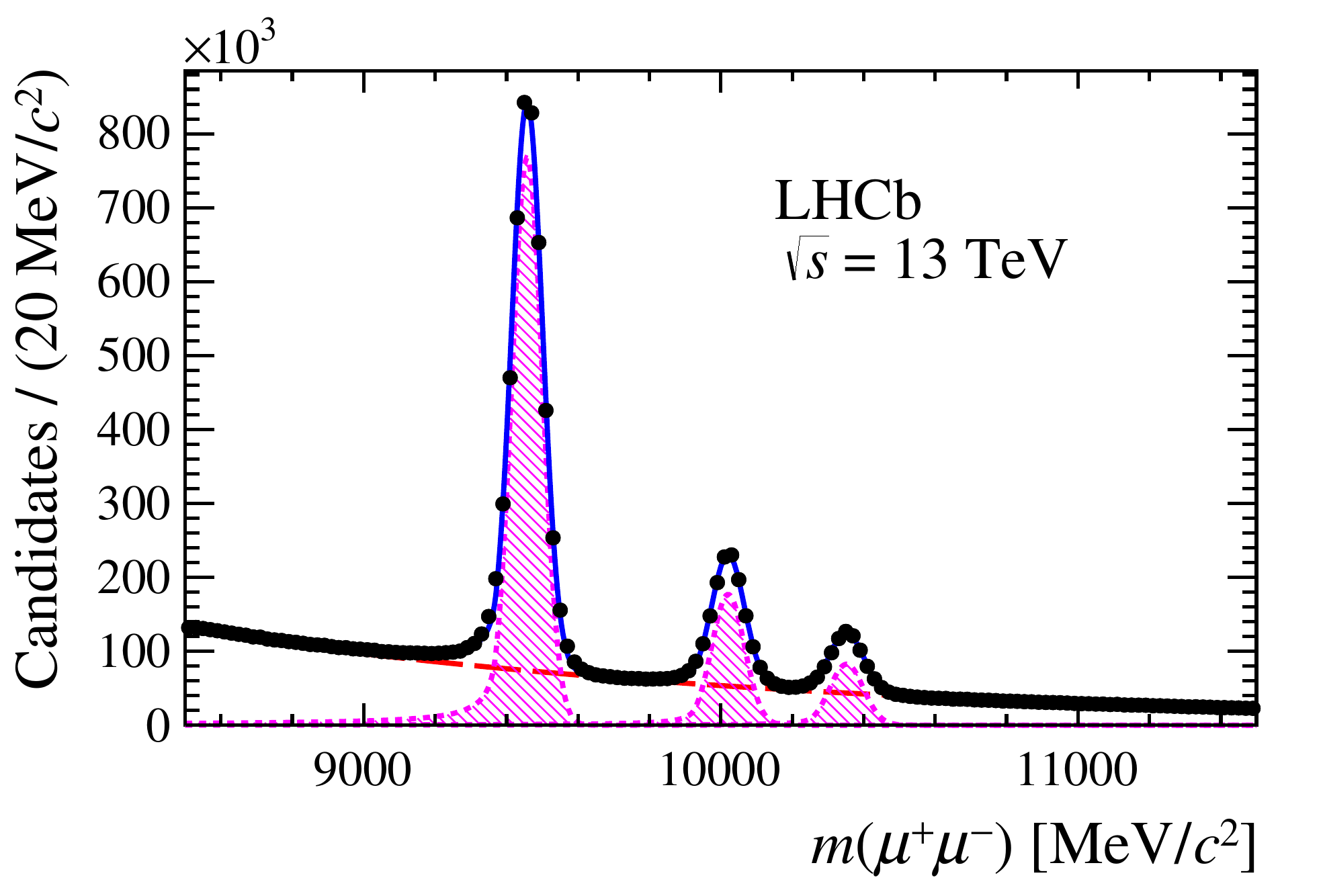}
    \put(-92,85){(c)}
	\includegraphics[width=0.49\textwidth]{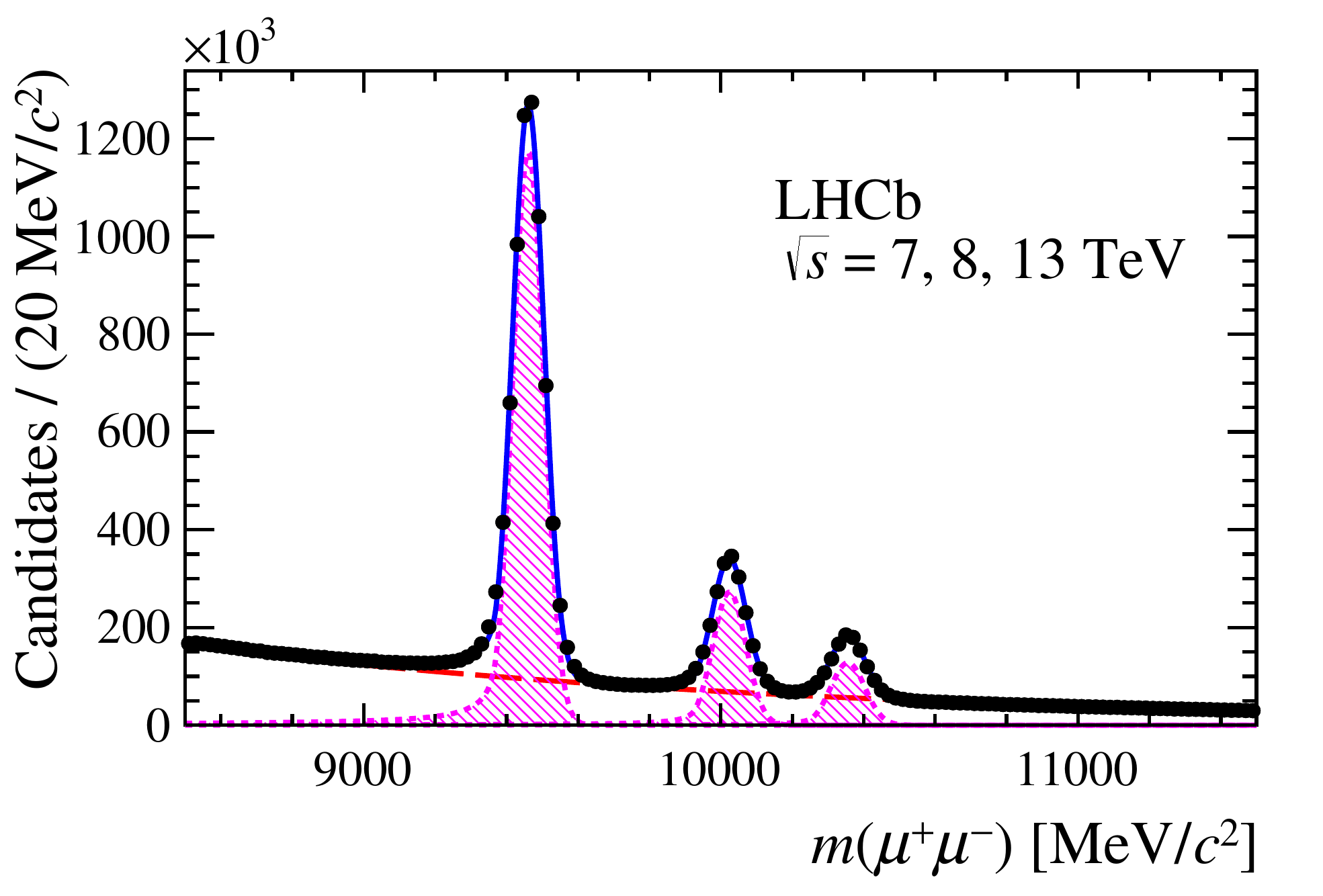}
    \put(-92,85){(d)}
    \caption{\label{fig:fitNorm}\small Distributions of $m(\mumu)$
	    for the normalisation datasets
        at $\proton\proton$ centre-of-mass energies of
	(a) 7\tevfix, (b) 8\tevfix, (c) 13\tevfix and (d) all combined.
    	The total fit function (solid blue line),
        the combinatorial background (dashed red line) and
        the $\OneS$, $\TwoS$ and $\ThreeS$ components (hatched magenta area)
        are shown overlaid.
    }
\end{figure}

\begin{figure}
	\centering
	\includegraphics[width=0.49\textwidth]{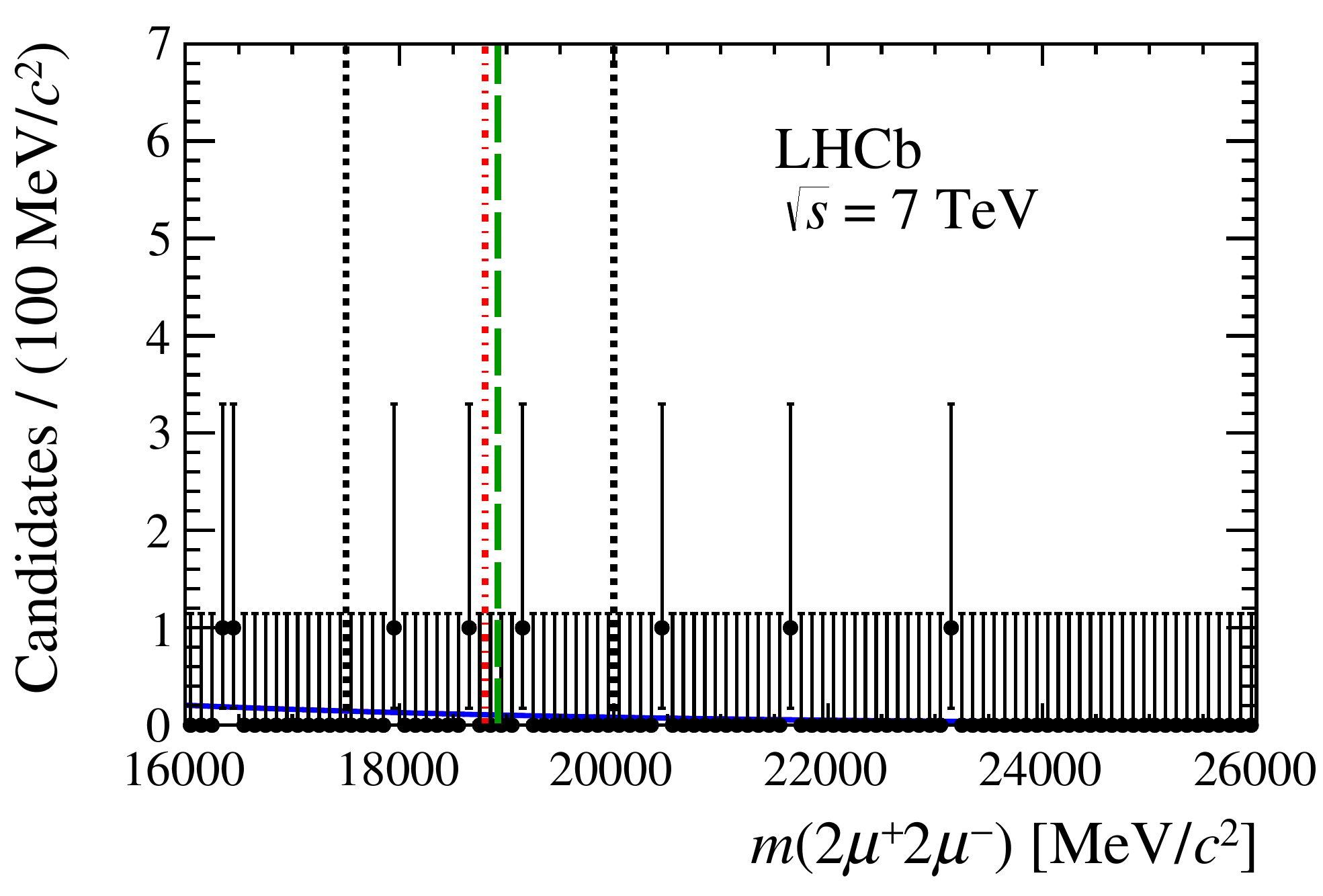}
    \put(-92,95){(a)}
	\includegraphics[width=0.49\textwidth]{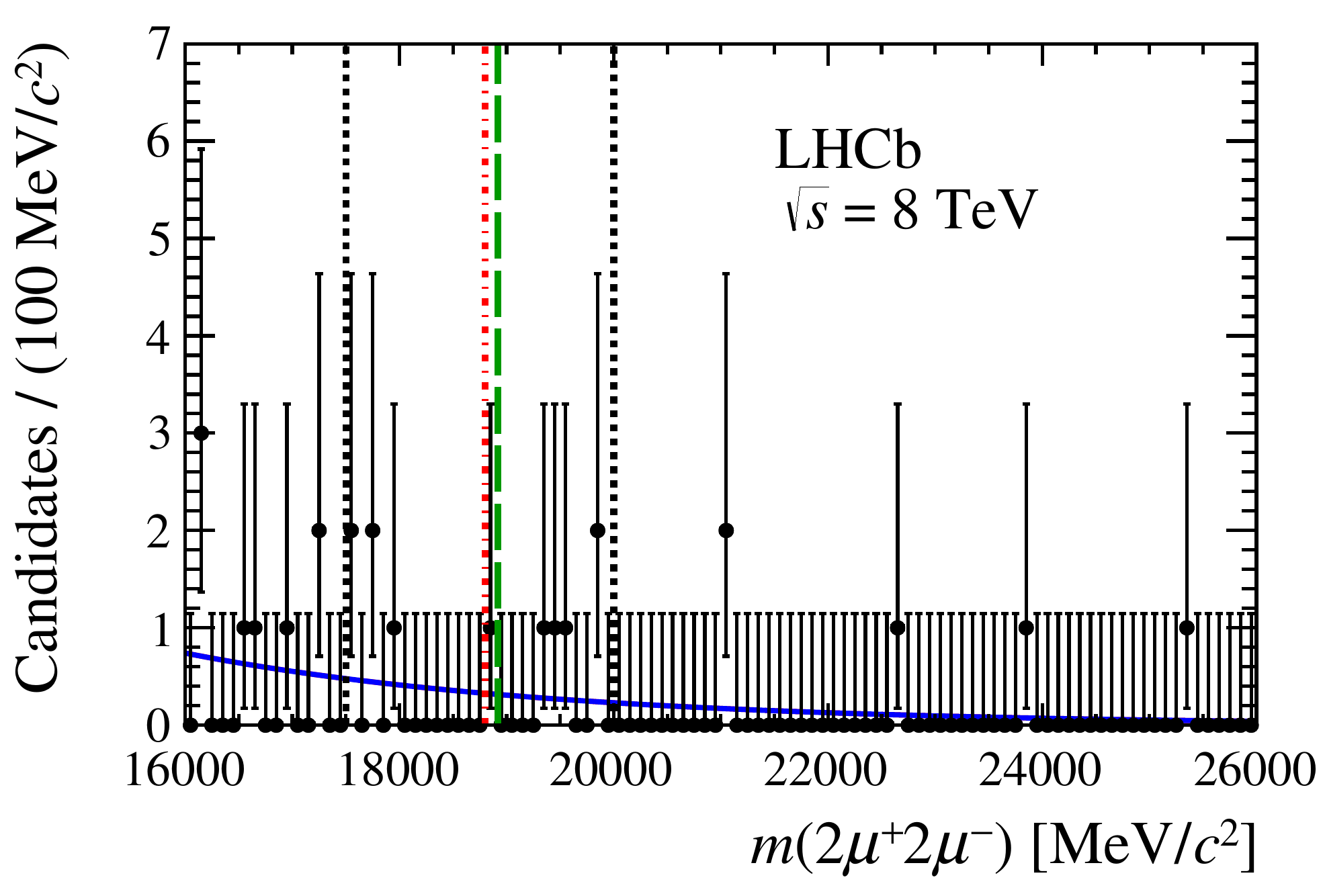}
    \put(-92,95){(b)}\\
	\includegraphics[width=0.49\textwidth]{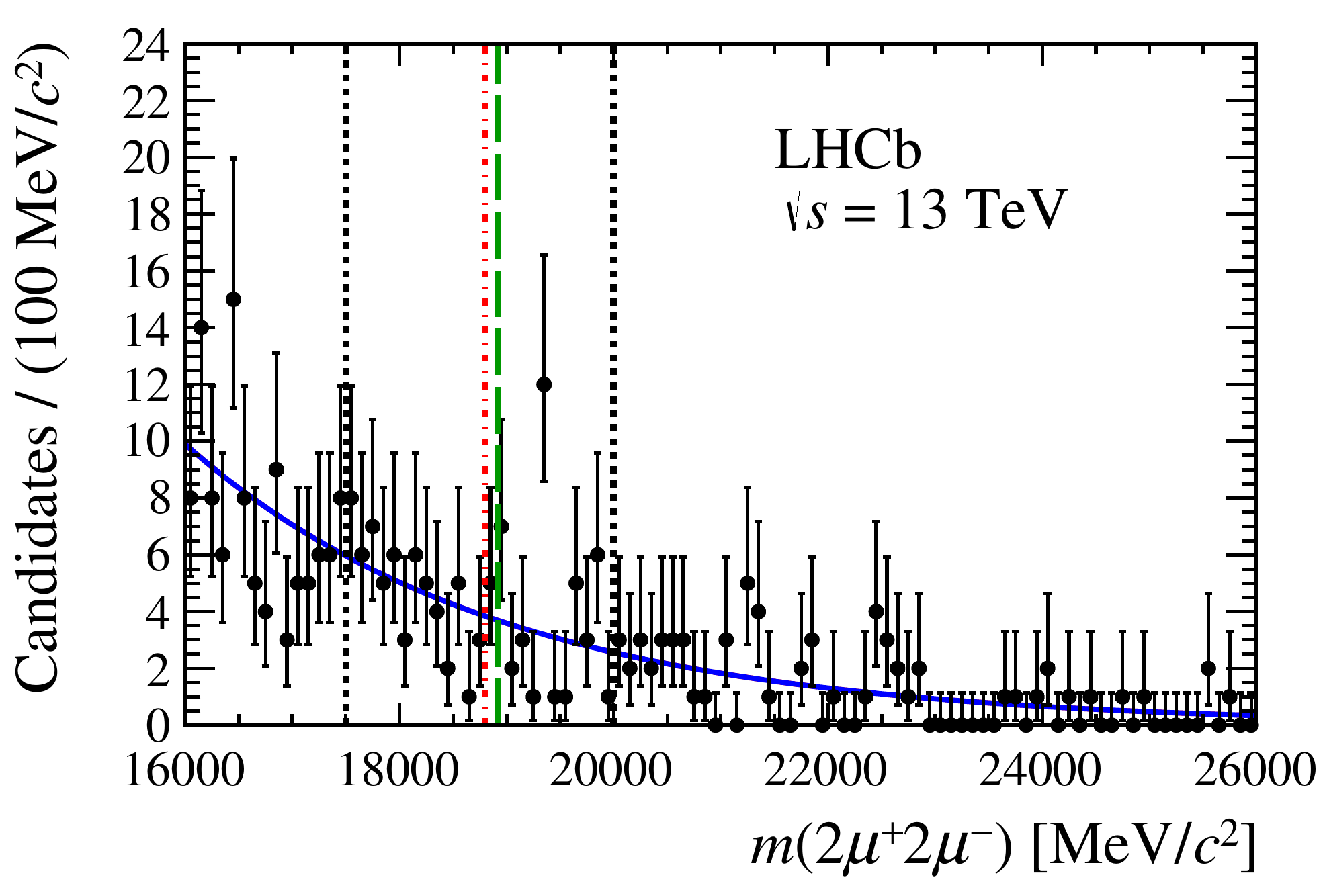}
    \put(-92,95){(c)}
	\includegraphics[width=0.49\textwidth]{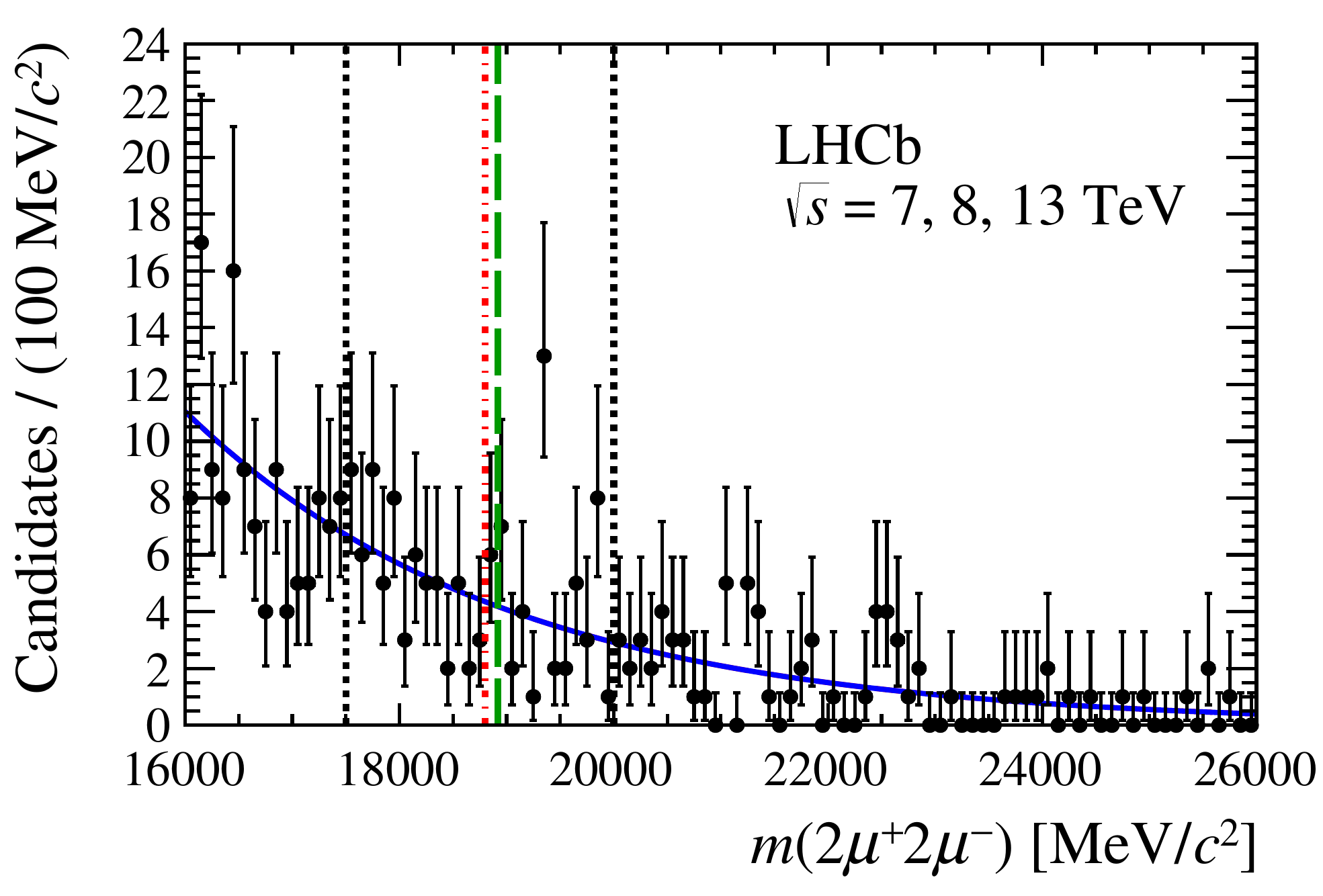}
    \put(-92,95){(d)}
    \caption{\label{fig:4mufit}\small Distributions of $m(2\mu^+2\mu^-)$
    for the signal datasets
        at $\proton\proton$ centre-of-mass energies of
	(a) 7\tevfix, (b) 8\tevfix, (c) 13\tevfix and (d) all combined,
    	using a bin size comparable to the expected
        $X$ mass resolution.
    	In each case the region around the corresponding
    	$\OneS$ peak has been selected.
    	The background-only fit function (solid blue line) is
        shown overlaid.
    	The  dotted black lines indicate the range in which limits are set on the product of the $X$ production cross-section and branching fractions. The dash-dotted red and long-dashed green
    lines show the positions of the $\eta_b\eta_b$ and $\OneS\OneS$ thresholds, respectively.
    }
\end{figure}

The fits to the $m(\mumu)$ distributions in
the normalisation datasets are shown in Fig.~\ref{fig:fitNorm}.
The fitted $\OneS$ yields in the range $R_{\OneS}\equiv\mu_{\OneS}\pm2.5\sigma_{\OneS}$ are $(0.694\pm0.012)\times10^6$, $(1.562\pm0.028)\times10^6$,
$(4.11\pm0.08)\times10^6$ and $(6.37\pm 0.12)\times10^6$ for the 7, 8, 13\tev
and combined datasets, respectively. The uncertainties include systematic components
due to the choice of shapes to describe the signal and background components.
Only candidates in the signal dataset with $m(\mup\mun)$ in the range
$R_{\OneS}$ are retained for the fits
to the distributions of $m(2\mup2\mun)$, which includes a small fraction of non-$\OneS$ background.
Background-only fits to the signal datasets are shown in Fig.~\ref{fig:4mufit}.
No significant signal excess is observed.
The largest
deviation occurs at a mass of approximately 19.35\gevcc, above the $\eta_b\eta_b$
and $\OneS\OneS$ thresholds, with a local significance of $2.5$
standard deviations.

\section{Normalisation factor}
\label{sec:normfactor}

Upper limits are set for different
$X$ mass hypotheses on the quantity
\begin{equation}
S \equiv \sigma(\proton\proton\to X)\times\mathcal{B}(X\to\OneS\mumu) \times \mathcal{B}(\OneS\to\mu^+\mu^-),
\end{equation}
where $\sigma(\proton\proton\to X)$ is the $X$
production cross-section, and $\mathcal{B}(X\to\OneS\mumu)$
and $\mathcal{B}(\OneS\to\mu^+\mu^-)$ are the branching fractions of the
$X\to\OneS\mumu$ and ${\OneS\to\mu^+\mu^-}$ decays, respectively.
To set limits on $S$,
the signal yield is parameterised as $N_{\rm sig} = S/f_{\rm norm}$
with
\begin{equation}
	f_{\rm norm} = \frac{\sigma(\proton\proton\to\OneS)\times\mathcal{B}(\OneS\to\mumu)}{N_{\rm norm}} \times \frac{\epsilon_{\rm norm}}{\epsilon_{\rm sig}}\,,
\end{equation}
where $\sigma(\proton\proton\to\OneS)$ is
the production cross-section of the $\OneS$ meson~\cite{LHCb-PAPER-2015-045,LHCb-PAPER-2018-002}
within the same fiducial volume as the signal.
The $\OneS\to\mumu$ yield within the range $R_{\OneS}$
is given by $N_{\rm norm}$, and
$\epsilon_{\rm sig (norm)}$ is the efficiency with which the signal (normalisation) channel is triggered, reconstructed and selected.

The relative efficiency of the reconstruction and selection requirements placed on
the corresponding signal and normalisation datasets is defined as
\begin{equation}
	\frac{\epsilon_{\rm sig}}{\epsilon_{\rm norm}} = \frac{\epsilon^{\rm geom}_{\rm sig}}{\epsilon^{\rm geom}_{\rm norm}} \times
							    \frac{\epsilon^{\rm sel}_{\rm sig}}{\epsilon^{\rm sel}_{\rm norm}} \times
							    \epsilon^{\rm PID}_{\rm sig} \times f^{\rm trk}_{\rm sig}\,,
\end{equation}
where $\epsilon^{\rm geom}$ is the efficiency with which the products of the $X$ or $\OneS$ decay all enter the LHCb geometric acceptance;
$\epsilon^{\rm sel}$ is the efficiency of the reconstruction and selection of $X$ or $\OneS$ candidates
within the geometric acceptance;
$\epsilon^{\rm PID}_{\rm sig}$ is the efficiency of the PID requirements placed on the additional muons in the signal decay;
and $f^{\rm trk}_{\rm sig}$ accounts for differences between data
and simulation in the tracking efficiency of the additional muons.
The geometric and selection efficiencies are determined from simulated samples, while the PID efficiency is determined from calibration data samples.
The ratio of efficiencies between the signal and normalisation samples
is determined to be $31.7\pm0.6\,\%$ ($35.2\pm1.2\,\%$) for the
$7,8\tev$ ($13\tev$) dataset, where the same efficiency is used for
7 and 8\tev collisions due to the similar performance of the LHCb
detector during these operational periods.

Uncertainties on these quantities give rise to systematic uncertainties in the fits to
the signal datasets and enter these fits as a Gaussian function constraining the value of
$f_{\rm norm}$. These systematic uncertainties are detailed further in Sec.~\ref{sec:syst}.
In the case of the combined dataset,
averages of the efficiency ratio and
normalisation cross-section,
weighted by the integrated luminosity of each subset,
are used to calculate $f_{\rm norm}$.
The values of $f_{\rm norm}$ are $11.1\pm1.5$, $6.49\pm0.25$, $3.27\pm0.24$ and $1.82\pm0.10\fb$
for the $7$, $8$, $13\tev$ and combined datasets, respectively.

\section{Systematic uncertainties}
\label{sec:syst}

Systematic uncertainties are included in the fits to the distribution of $m(2\mup2\mun)$ through additional Gaussian terms in the likelihood function
that constrain the values of four nuisance parameters:
$f_{\rm norm}$, $\sigma_{\OneS}$, $p_0$ and $p_1$.
Uncertainties on the normalisation yields, the $\OneS$ production cross-sections, and the relative efficiencies of the signal and normalisation channels
all contribute to the uncertainty on the $f_{\rm norm}$ parameter.
The uncertainty on $\sigma_{\OneS}$ is obtained from the fit to the $m(\mup\mun)$ distribution of the normalisation channel.
The linear coefficients of the $X$-mass-dependent resolution scale term are constrained
according to the uncertainties on these parameters from fits to simulated data.

The relative uncertainties on the $\sigma_{\OneS}$, $p_0$ and $p_1$ parameters are $\lesssim0.1\,\%$, $0.5\,\%$ and $46\,\%$, respectively. Since
these parameters are weakly correlated with the signal yield their effects on the measured cross-section upper limits are negligible.
The uncertainty on the $f_{\rm norm}$ parameter for each dataset is dominated by uncertainties on the
normalisation cross-section ($2.8$ to $6.3\,\%$) and the tracking efficiency correction ($0.8$ to $3.1\,\%$). The systematic uncertainties
from efficiencies related to particle identification or
geometrical acceptance are at the level of 1.0\,\% or less.
For the $7\tev$ result, a discrepancy is observed in the efficiency- and cross-section-corrected $\OneS$ yield relative to the other datasets. An additional  uncertainty of $13.5\,\%$ is assigned to account for this.
This uncertainty increases the limits on the cross section at 7 TeV by $<4\,\%$ and has no effect on the quoted combined limits.
The limits reported on the $X$ production cross-section are all statistically dominated.

\section{Limit setting}

\begin{figure}
	\centering
	\includegraphics[width=0.49\textwidth]{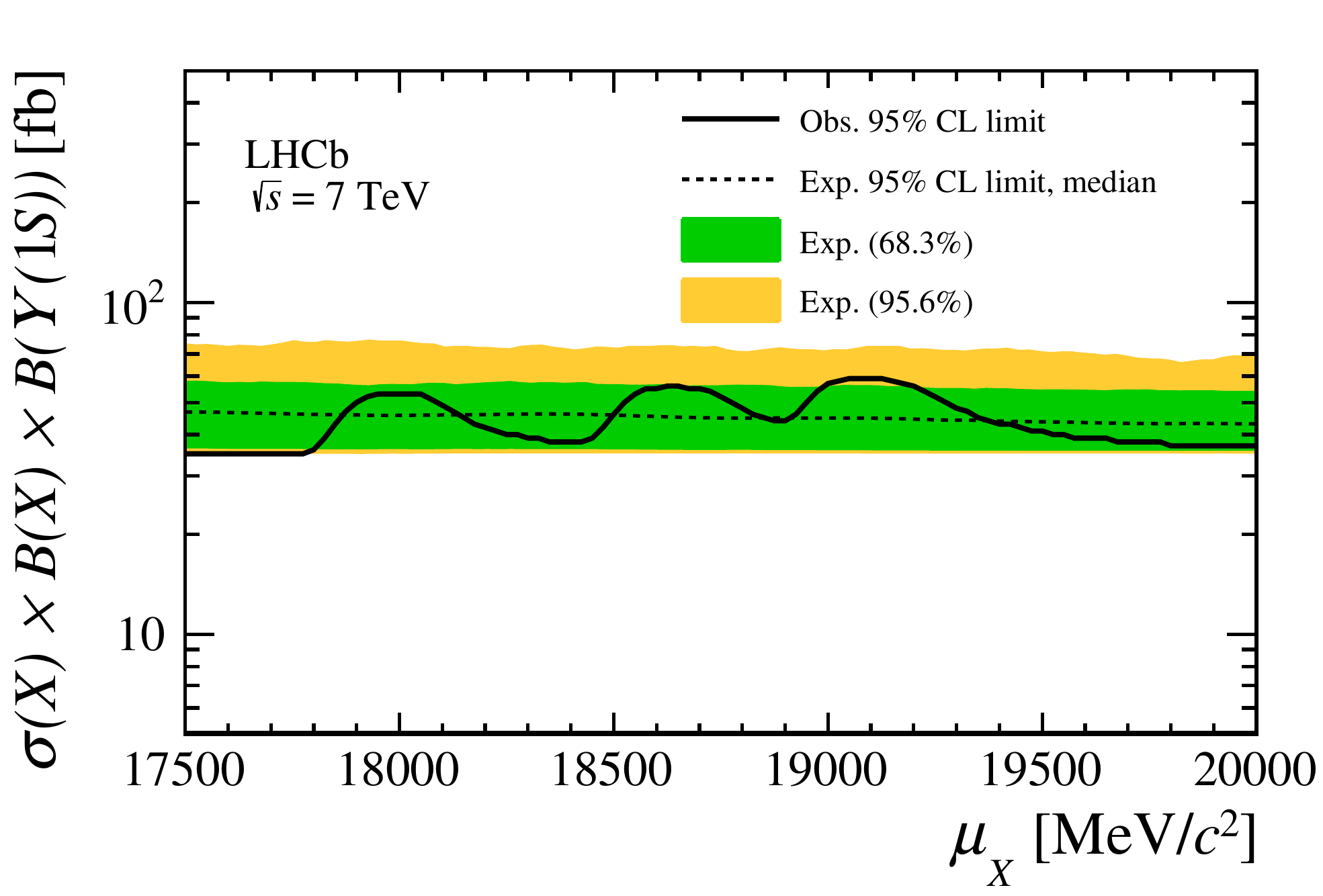}
    \put(-180,35){(a)}
	\includegraphics[width=0.49\textwidth]{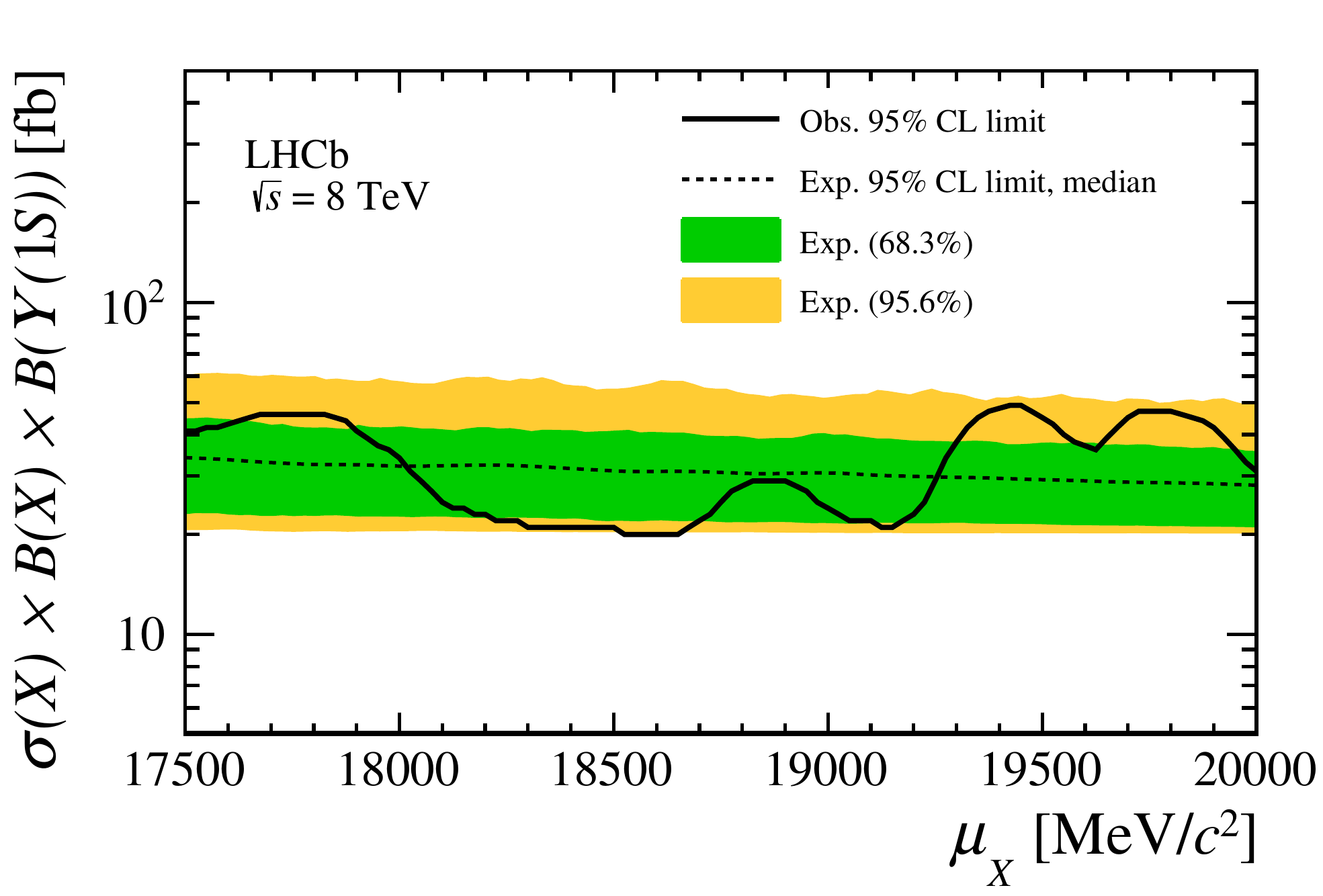}
    \put(-180,35){(b)}\\
	\includegraphics[width=0.49\textwidth]{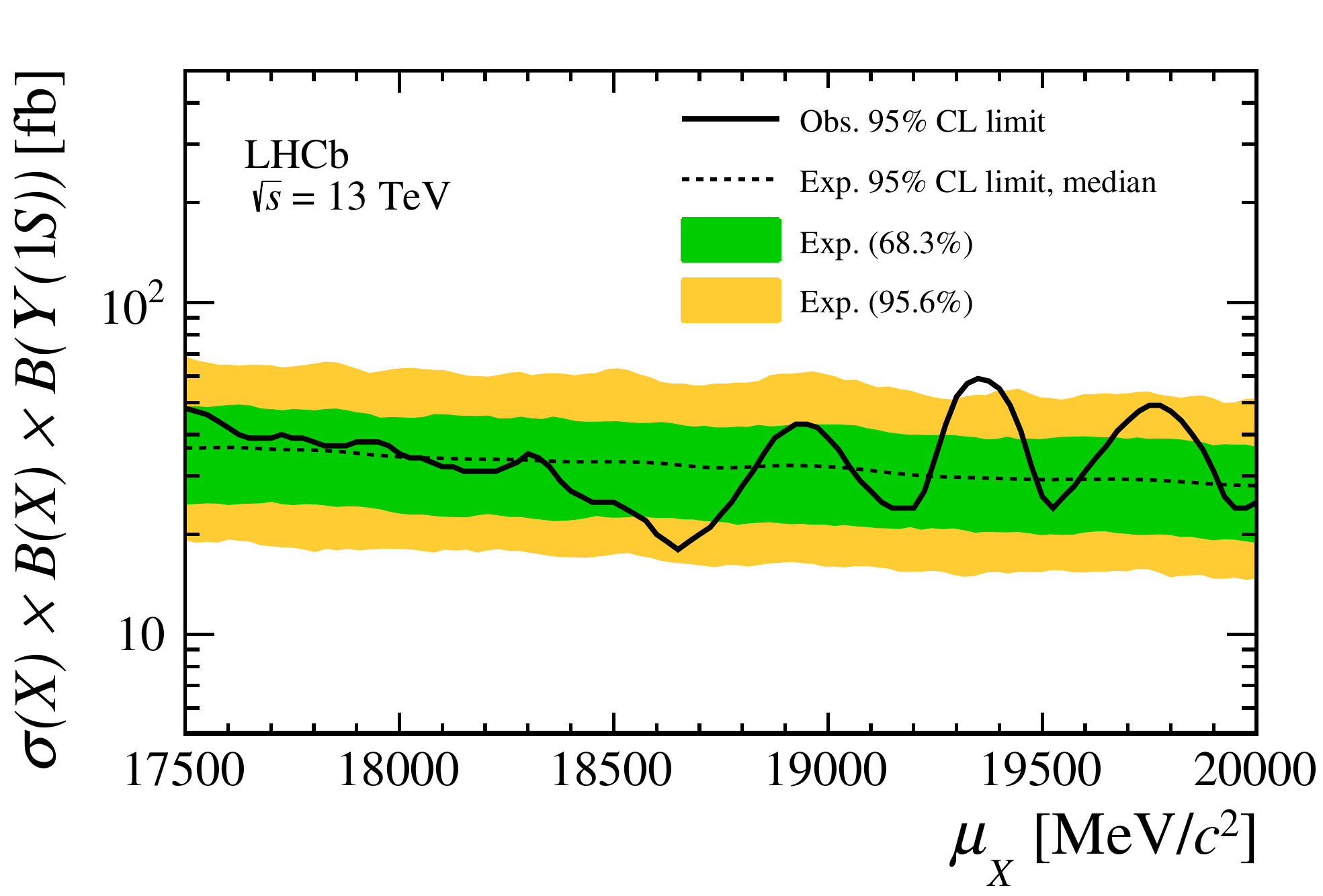}
    \put(-180,35){(c)}
	\includegraphics[width=0.49\textwidth]{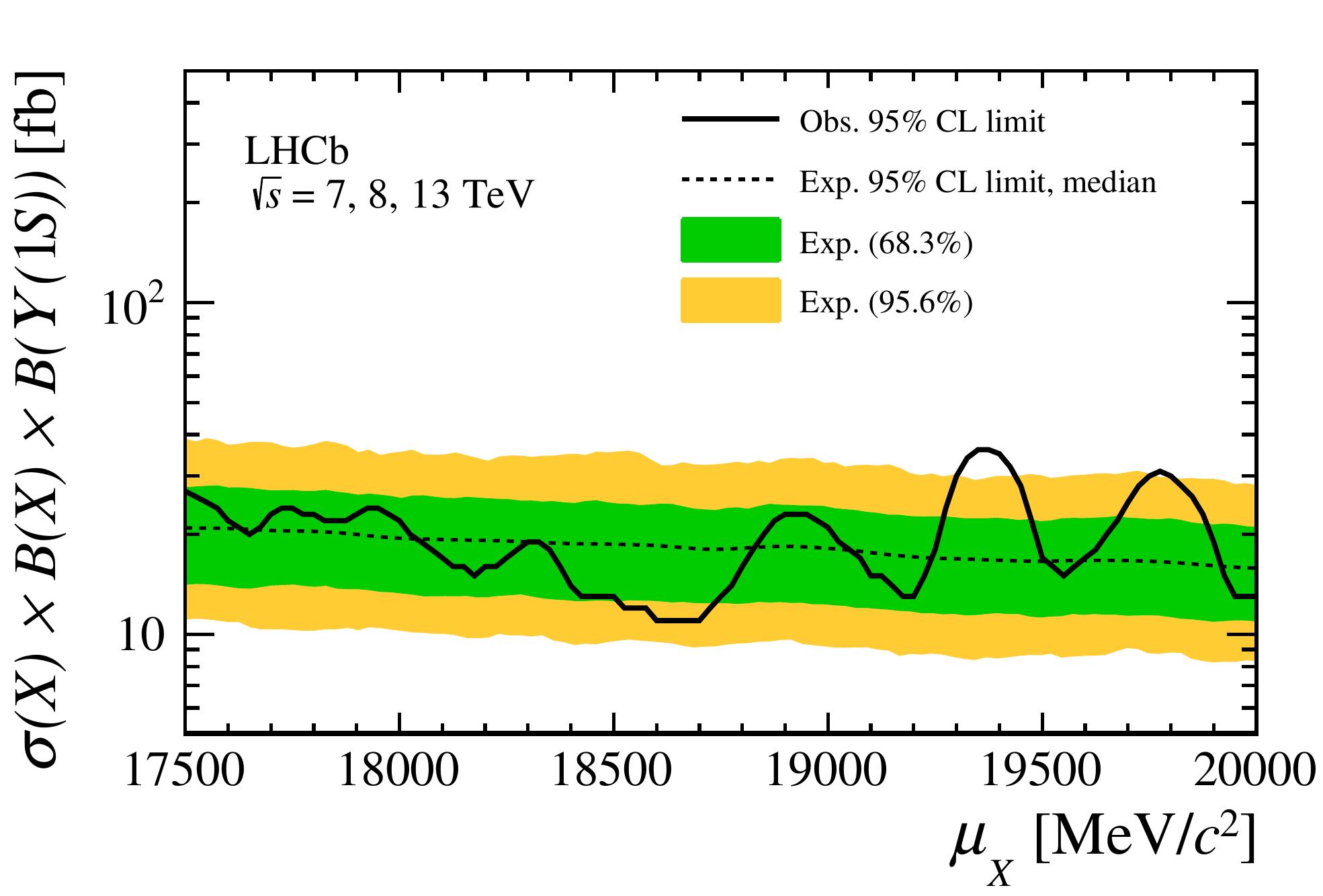}
    \put(-180,35){(d)}
    \caption{\label{fig:massScanLimitsBrazil}\small
        The 95\,\% CL upper limits on $S\equiv \sigma(\proton\proton\to X)\times\mathcal{B}(X\to\OneS\mumu) \times \mathcal{B}(\OneS\to\mu^+\mu^-)$
        as functions of the $X$ mass hypothesis at $\proton\proton$ centre-of-mass energies of (a) 7\tevfix, (b)~8\tevfix and (c) 13\tevfix
        and (d) all combined.
            }
\end{figure}

For each signal dataset, upper limits are set on $S$
as functions of the $X$ mass, $\mu_{X}$, in the range $[17.5,20.0]\gevcc$ using the
following procedure.
For each fixed $X$ mass, the likelihood profile as a function of $S$ is integrated to determine upper limits
on the cross-section at 90\,\% and 95\,\% confidence levels (CL).
This procedure is applied at each of 101 values of the $X$ mass.
The 90\,\% and 95\,\% CL limits are
tabulated in the \hyperref[sec:appendix]{Appendix}.
Background-only pseudoexperiments are generated
at each scan point to determine the
expected 95\,\% CL upper limit and corresponding one and two standard
deviation intervals, as shown in Fig.~\ref{fig:massScanLimitsBrazil}.
No significant excess is seen at any mass hypothesis for any dataset.

The analysis is repeated with only a single candidate decay retained
for each event
(chosen at random), with a more stringent requirement on the pseudorapidity of
the muons as was previously used in Ref.~\cite{LHCB-PAPER-2017-028}.
In addition, the effect of the assumption that the $X$ decays according to a phase-space distribution is tested by evaluating the efficiency for both
$m(\mup\mun)$ less than $2\gevcc$ and $m(\mup\mun)$ greater than $7\gevcc$ for the muon pairs that do not come from the \OneS\ decay. The efficiency varies $\pm24\,\%$ with respect to the total efficiency under the assumption of a phase-space decay.
Finally, the limits are evaluated using different ranges around the $\OneS$ mass to select the signal dataset, separately for each year of the $\sqs=13\tev$ dataset, and for the $7$ and $8\tev$ datasets combined.
No significant differences are observed in the limits determined in each of these cross-checks.

\section{Conclusions}

In conclusion, a search is performed for the decay of the beautiful
tetraquark, $X$, to
the $\OneS\mumu$ final state. No significant excess is seen for any mass hypothesis in the range $[17.5,20.0]\gevcc$.
Upper limits are set on the value of
${\sigma(\proton\proton\to X)\times\mathcal{B}(X\to\OneS\mumu) \times \mathcal{B}(\OneS\to\mu^+\mu^-)}$
at centre-of-mass energies ${\sqs=7\tev}$, $8\tev$ and $13\tev$ as functions of the $X$ mass hypothesis (see Appendix).
An upper limit is also set on the combined dataset
using the average of the $\OneS$ cross-section,
weighted by the integrated luminosity of each
subset, resulting in
upper limits of $\mathcal{O}(10\fb)$.
Improved sensitivity for this state will be obtained using data collected during future running periods of the LHC using an updated LHCb detector~\cite{LHCb-TDR-012,LHCb-PII-EoI,LHCb-PII-Physics}.

\section*{Acknowledgements}
\noindent We express our gratitude to our colleagues in the CERN
accelerator departments for the excellent performance of the LHC. We
thank the technical and administrative staff at the LHCb
institutes. We acknowledge support from CERN and from the national
agencies: CAPES, CNPq, FAPERJ and FINEP (Brazil); MOST and NSFC
(China); CNRS/IN2P3 (France); BMBF, DFG and MPG (Germany); INFN
(Italy); NWO (Netherlands); MNiSW and NCN (Poland); MEN/IFA
(Romania); MinES and FASO (Russia); MinECo (Spain); SNSF and SER
(Switzerland); NASU (Ukraine); STFC (United Kingdom); NSF (USA).  We
acknowledge the computing resources that are provided by CERN, IN2P3
(France), KIT and DESY (Germany), INFN (Italy), SURF (Netherlands),
PIC (Spain), GridPP (United Kingdom), RRCKI and Yandex
LLC (Russia), CSCS (Switzerland), IFIN-HH (Romania), CBPF (Brazil),
PL-GRID (Poland) and OSC (USA). We are indebted to the communities
behind the multiple open-source software packages on which we depend.
Individual groups or members have received support from AvH Foundation
(Germany), EPLANET, Marie Sk\l{}odowska-Curie Actions and ERC
(European Union), ANR, Labex P2IO and OCEVU, and R\'{e}gion
Auvergne-Rh\^{o}ne-Alpes (France), Key Research Program of Frontier
Sciences of CAS, CAS PIFI, and the Thousand Talents Program (China),
RFBR, RSF and Yandex LLC (Russia), GVA, XuntaGal and GENCAT (Spain),
Herchel Smith Fund, the Royal Society, the English-Speaking Union and
the Leverhulme Trust (United Kingdom).

\clearpage

{\noindent\normalfont\bfseries\Large Appendix}

\appendix
\label{sec:appendix}

\begin{figure}[h]
	\centering
	\includegraphics[width=0.49\textwidth]{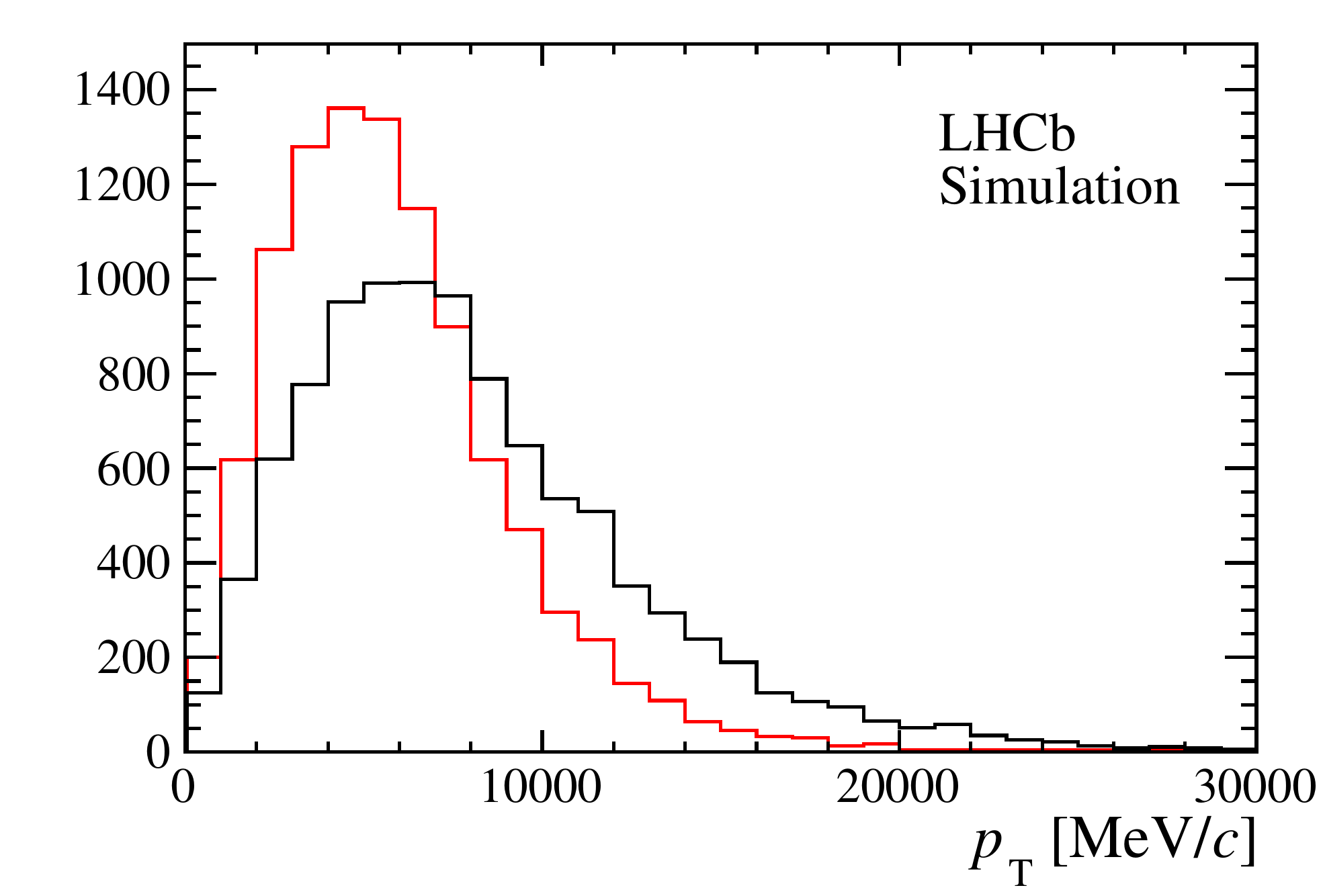}
    \put(-65,100){(a)}
    \put(-100,42){\includegraphics[width=0.17\textwidth]{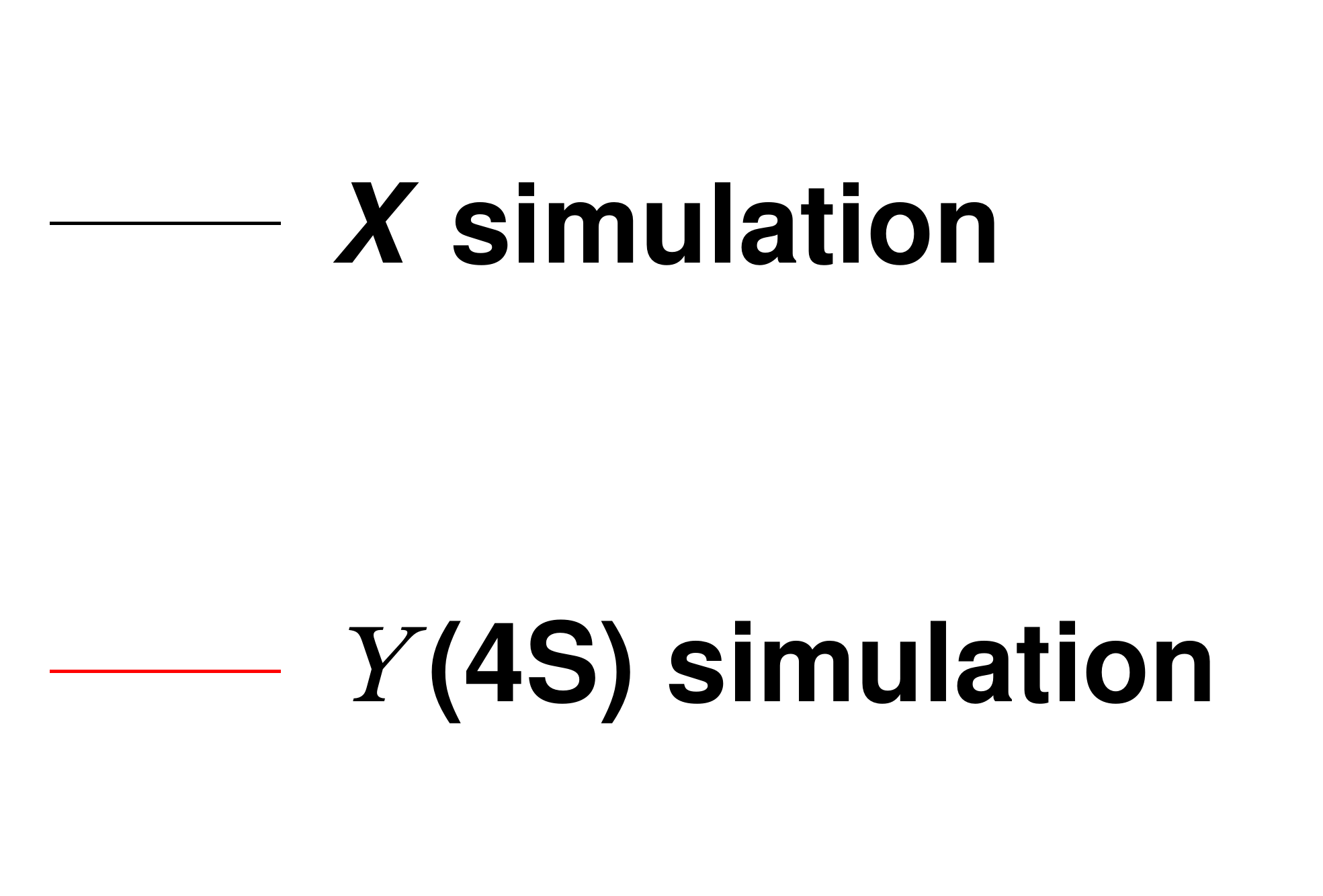}}
	\includegraphics[width=0.49\textwidth]{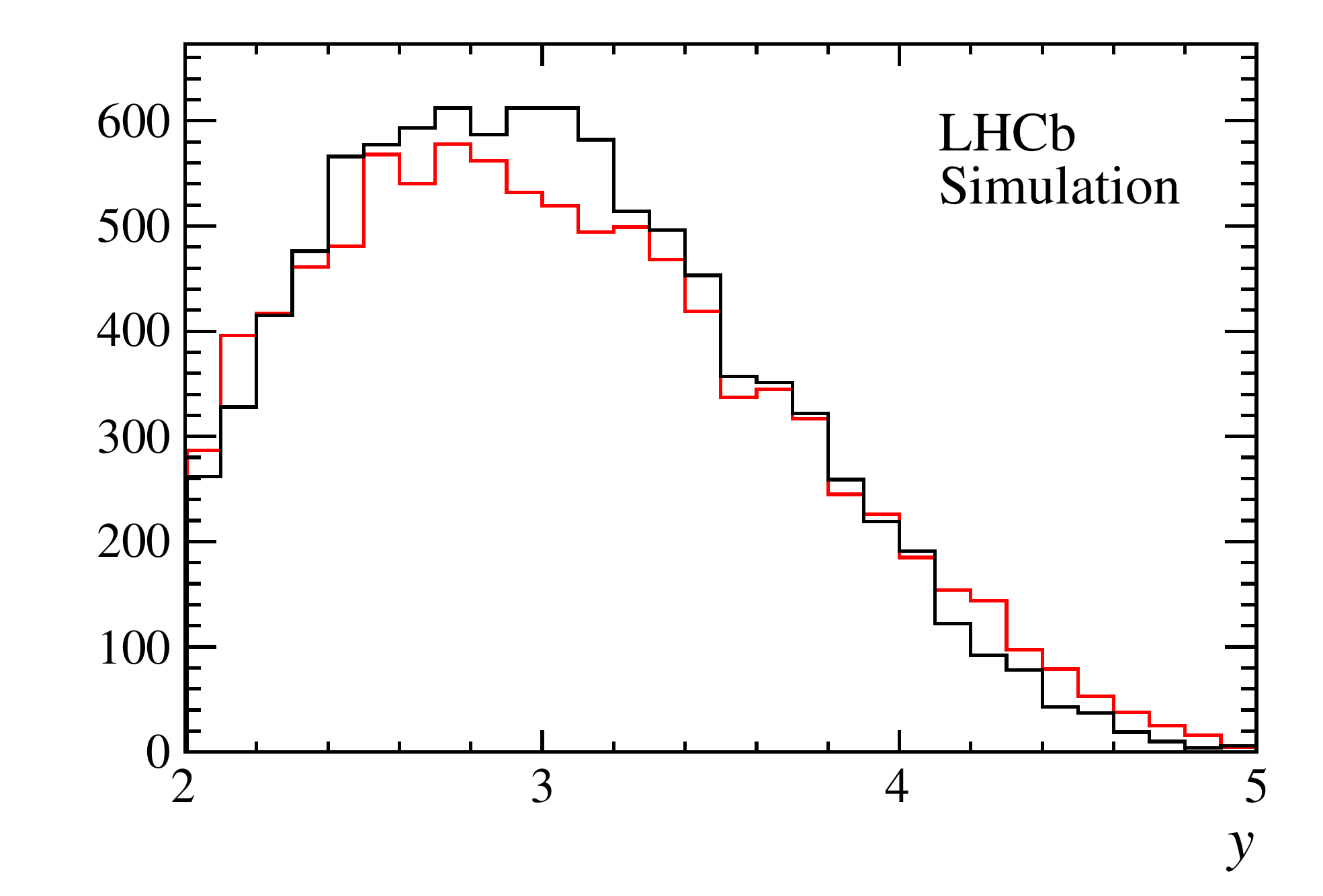}
    \put(-65,100){(b)}\\
	\includegraphics[width=0.49\textwidth]{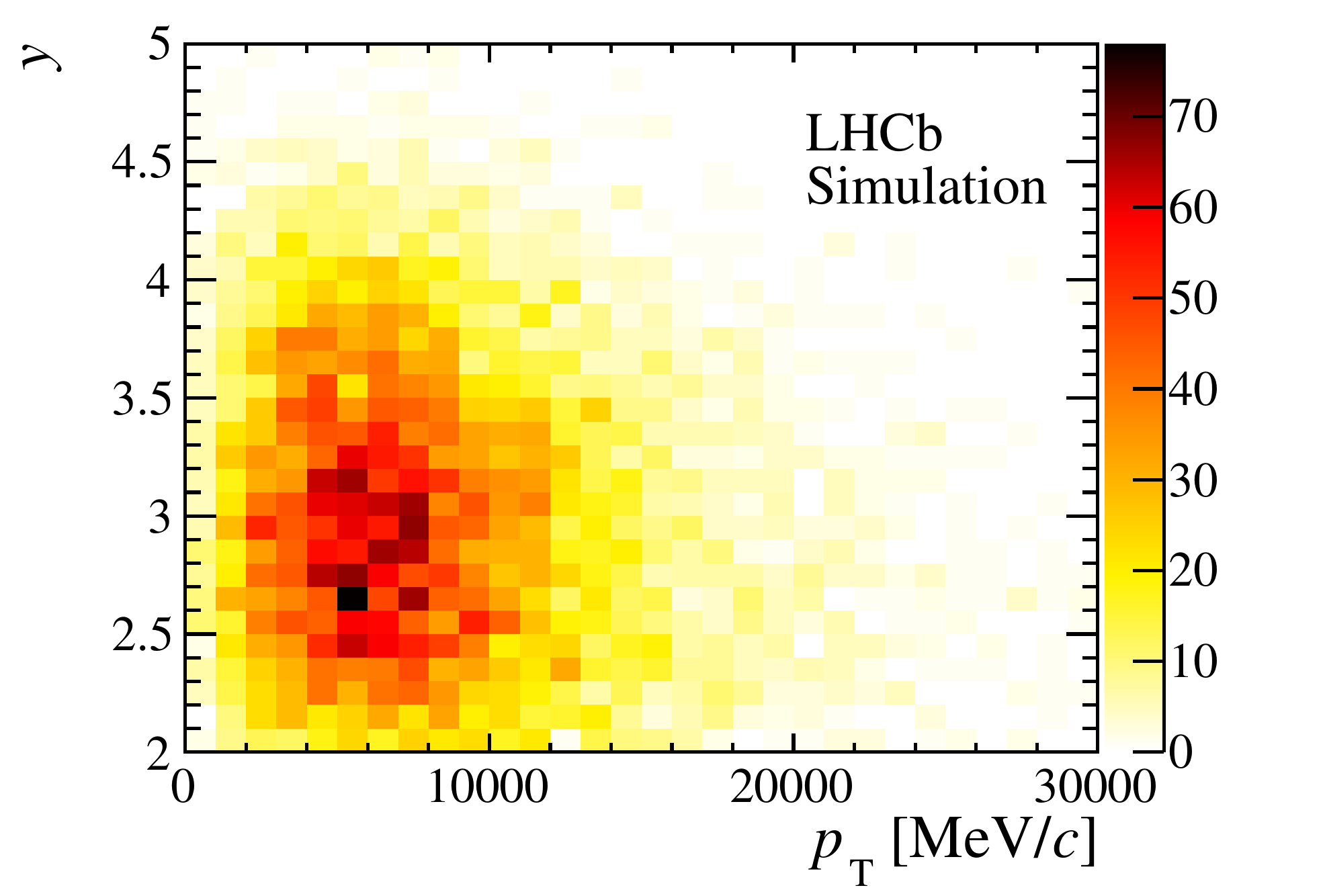}
    \put(-87,100){(c)}
    \caption{\label{fig:simkinematics}\small
        The kinematic distribution of (black) simulated $X$ particles in
        (a) $\pt$ and (b) rapidity, and (c) the 2D distribution.
        For comparison, the kinematic distribution of (red) simulated $\FourS$
        particles is also shown.
            }
\end{figure}

\begin{table}[h]
		\caption{\label{tab:massScanFits0}\small
		Upper limits on $\sigma(\proton\proton\to X) \times \mathcal{B}(X\to\OneS\mumu) \times \mathcal{B}(\OneS\to\mumu)$
		for different $X$ mass hypotheses in the range $[17.5,18.4]\gevcc$.
		}
        \begin{center}
		\begin{tabular}{c|ccc|c}
			Mass & \multicolumn{4}{c}{Upper limit 90\,\% (95\,\%) CL [\fb]} \\
			{[\mevcc]} & 7\tev & 8\tev & 13\tev & Combined \\
			\hline
 17500 & 27 (35) & 34 (41) & 41 (48) & 23 (27) \\
 17525 & 27 (35) & 34 (41) & 40 (47) & 23 (27) \\
 17550 & 27 (35) & 35 (42) & 39 (46) & 22 (26) \\
 17575 & 27 (35) & 35 (42) & 37 (44) & 20 (24) \\
 17600 & 27 (35) & 36 (43) & 35 (42) & 19 (23) \\
 17625 & 27 (35) & 36 (44) & 33 (40) & 17 (21) \\
 17650 & 27 (35) & 37 (45) & 32 (39) & 17 (21) \\
 17675 & 27 (35) & 38 (46) & 32 (39) & 18 (22) \\
 17700 & 27 (35) & 38 (46) & 33 (39) & 19 (23) \\
 17725 & 27 (35) & 38 (46) & 33 (40) & 20 (24) \\
 17750 & 27 (35) & 38 (46) & 33 (39) & 20 (24) \\
 17775 & 27 (35) & 38 (46) & 32 (39) & 20 (24) \\
 17800 & 28 (36) & 39 (46) & 32 (38) & 19 (23) \\
 17825 & 29 (39) & 38 (46) & 31 (37) & 19 (22) \\
 17850 & 33 (43) & 38 (45) & 30 (37) & 18 (22) \\
 17875 & 37 (47) & 36 (44) & 30 (37) & 18 (22) \\
 17900 & 40 (50) & 34 (41) & 31 (38) & 19 (23) \\
 17925 & 42 (52) & 32 (39) & 32 (38) & 20 (24) \\
 17950 & 43 (53) & 30 (37) & 31 (38) & 20 (24) \\
 17975 & 43 (53) & 29 (36) & 30 (37) & 19 (23) \\
 18000 & 43 (53) & 27 (34) & 29 (35) & 18 (22)\\
 18025 & 43 (53) & 25 (31) & 28 (34) & 17 (21)\\
 18050 & 42 (53) & 22 (29) & 27 (34) & 16 (19)\\
 18075 & 41 (51) & 21 (27) & 27 (33) & 15 (18)\\
 18100 & 39 (49) & 20 (25) & 26 (32) & 14 (17)\\
 18125 & 37 (47) & 19 (24) & 26 (32) & 13 (16)\\
 18150 & 35 (45) & 18 (24) & 25 (31) & 13 (16)\\
 18175 & 34 (43) & 18 (23) & 25 (31) & 13 (16)\\
 18200 & 33 (42) & 17 (23) & 25 (31) & 13 (16)\\
 18225 & 32 (41) & 17 (22) & 25 (31) & 13 (17)\\
 18250 & 31 (40) & 17 (22) & 26 (32) & 14 (18)\\
 18275 & 30 (40) & 17 (22) & 28 (33) & 15 (19)\\
 18300 & 30 (39) & 16 (21) & 29 (35) & 16 (19)\\
 18325 & 30 (39) & 16 (21) & 28 (34) & 16 (19)\\
 18350 & 29 (38) & 16 (21) & 27 (32) & 15 (18)\\
 18375 & 29 (38) & 16 (21) & 24 (29) & 13 (16)\\
	\end{tabular}
	\end{center}
\end{table}

\begin{table}[]
		\caption{\label{tab:massScanFits1}\small
		Upper limits on $\sigma(\proton\proton\to X) \times \mathcal{B}(X\to\OneS\mumu) \times \mathcal{B}(\OneS\to\mumu)$
		for different $X$ mass hypotheses in the range $[18.4,19.3]\gevcc$.
		}
        \begin{center}
		\begin{tabular}{c|ccc|c}
			Mass & \multicolumn{4}{c}{Upper limit 90\,\% (95\,\%) CL (\fb)} \\
			{[\mevcc]} & 7\tev & 8\tev & 13\tev & Combined \\
			\hline
 18400 & 29 (38) & 16 (21) & 22 (27) & 11 (14)\\
 18425 & 29 (38) & 16 (21) & 20 (26) & 11 (13)\\
 18450 & 30 (39) & 16 (21) & 20 (25) & 10 (13)\\
 18475 & 32 (42) & 16 (21) & 20 (25) & 10 (13)\\
 18500 & 36 (46) & 16 (21) & 20 (25) & 10 (13)\\
 18525 & 40 (50) & 16 (20) & 19 (24) & 10 (13)\\
 18550 & 43 (53) & 16 (20) & 18 (23) & 10 (12)\\
 18575 & 44 (55) & 15 (20) & 17 (22) & \phantom{0}9 (12)\\
 18600 & 45 (55) & 15 (20) & 16 (20) & \phantom{0}9 (12)\\
 18625 & 45 (56) & 15 (20) & 15 (19) & \phantom{0}9 (11)\\
 18650 & 45 (56) & 16 (20) & 14 (18) & \phantom{0}8 (11)\\
 18675 & 45 (55) & 16 (21) & 14 (19) & \phantom{0}8 (11)\\
 18700 & 44 (55) & 17 (22) & 15 (20) & \phantom{0}9 (11)\\
 18725 & 43 (54) & 18 (23) & 17 (21) & \phantom{0}9 (12)\\
 18750 & 42 (52) & 20 (25) & 18 (23) & 10 (13)\\
 18775 & 40 (50) & 21 (27) & 20 (25) & 11 (14)\\
 18800 & 38 (48) & 22 (28) & 23 (28) & 13 (16)\\
 18825 & 36 (46) & 23 (29) & 26 (31) & 15 (18)\\
 18850 & 35 (45) & 23 (29) & 29 (35) & 17 (20)\\
 18875 & 34 (44) & 23 (29) & 32 (39) & 18 (22)\\
 18900 & 34 (44) & 23 (29) & 35 (41) & 20 (23)\\
 18925 & 35 (46) & 22 (28) & 36 (43) & 20 (24)\\
 18950 & 39 (50) & 21 (27) & 37 (43) & 20 (23)\\
 18975 & 43 (54) & 19 (25) & 35 (42) & 19 (22)\\
 19000 & 46 (57) & 18 (24) & 33 (39) & 18 (21)\\
 19025 & 47 (58) & 18 (23) & 30 (36) & 16 (20)\\
 19050 & 48 (59) & 17 (22) & 26 (32) & 15 (18)\\
 19075 & 48 (59) & 17 (22) & 24 (29) & 14 (17)\\
 19100 & 48 (59) & 16 (22) & 22 (27) & 13 (16)\\
 19125 & 48 (59) & 16 (21) & 20 (25) & 12 (15)\\
 19150 & 48 (58) & 16 (21) & 19 (24) & 11 (14)\\
 19175 & 47 (57) & 16 (22) & 19 (24) & 11 (14)\\
 19200 & 46 (56) & 17 (23) & 19 (24) & 11 (14)\\
 19225 & 44 (54) & 19 (25) & 22 (27) & 12 (15)\\
 19250 & 41 (52) & 23 (29) & 27 (34) & 15 (19)\\
 19275 & 39 (50) & 27 (34) & 36 (43) & 21 (25)\\
	\end{tabular}
	\end{center}
\end{table}

\begin{table}[]
		\caption{\label{tab:massScanFits4}\small
		Upper limits on $\sigma(\proton\proton\to X) \times \mathcal{B}(X\to\OneS\mumu) \times \mathcal{B}(\OneS\to\mumu)$
		for different $X$ mass hypotheses in the range $[19.3,20.0]\gevcc$.
		}
        \begin{center}
		\begin{tabular}{c|ccc|c}
			Mass & \multicolumn{4}{c}{Upper limit 90\,\% (95\,\%) CL (\fb)} \\
			{[\mevcc]} & 7\tev & 8\tev & 13\tev & Combined \\
			\hline
 19300 & 38 (48) & 31 (38) & 45 (52) & 27 (31)\\
 19325 & 36 (47) & 34 (42) & 50 (57) & 31 (35)\\
 19350 & 35 (45) & 37 (45) & 52 (59) & 32 (36)\\
 19375 & 34 (44) & 40 (47) & 51 (58) & 32 (36)\\
 19400 & 33 (43) & 41 (48) & 48 (55) & 31 (35)\\
 19425 & 33 (43) & 42 (49) & 42 (49) & 29 (32)\\
 19450 & 32 (42) & 41 (49) & 34 (41) & 24 (28)\\
 19475 & 32 (41) & 40 (47) & 26 (32) & 19 (22)\\
 19500 & 31 (41) & 38 (45) & 21 (26) & 14 (18)\\
 19525 & 31 (40) & 36 (43) & 19 (24) & 13 (16)\\
 19550 & 31 (40) & 33 (40) & 20 (26) & 13 (16)\\
 19575 & 30 (39) & 31 (38) & 22 (28) & 13 (16)\\
 19600 & 30 (39) & 29 (37) & 25 (31) & 14 (17)\\
 19625 & 30 (39) & 29 (36) & 28 (34) & 15 (19)\\
 19650 & 29 (39) & 31 (39) & 31 (37) & 17 (21)\\
 19675 & 29 (38) & 34 (42) & 34 (41) & 19 (23)\\
 19700 & 29 (38) & 37 (45) & 38 (44) & 22 (26)\\
 19725 & 29 (38) & 39 (47) & 41 (47) & 25 (29)\\
 19750 & 29 (38) & 40 (47) & 42 (49) & 27 (30)\\
 19775 & 29 (38) & 40 (47) & 42 (49) & 27 (31)\\
 19800 & 29 (37) & 39 (47) & 41 (47) & 26 (30)\\
 19825 & 28 (37) & 39 (46) & 38 (44) & 25 (28)\\
 19850 & 28 (37) & 38 (45) & 34 (40) & 22 (26)\\
 19875 & 28 (37) & 37 (44) & 30 (36) & 19 (23)\\
 19900 & 28 (37) & 35 (42) & 25 (31) & 16 (19)\\
 19925 & 28 (37) & 32 (39) & 21 (26) & 12 (16)\\
 19950 & 28 (37) & 29 (36) & 19 (24) & 11 (13)\\
 19975 & 28 (37) & 26 (33) & 19 (24) & 10 (13)\\
 20000 & 28 (37) & 24 (31) & 20 (25) & 11 (13)\\
		\end{tabular}
	\end{center}
\end{table}

\clearpage

\addcontentsline{toc}{section}{References}
\ifx\mcitethebibliography\mciteundefinedmacro
\PackageError{LHCb.bst}{mciteplus.sty has not been loaded}
{This bibstyle requires the use of the mciteplus package.}\fi
\providecommand{\href}[2]{#2}

\newpage

\newpage

\centerline{\large\bf LHCb collaboration}
\begin{flushleft}
\small
R.~Aaij$^{27}$,
B.~Adeva$^{41}$,
M.~Adinolfi$^{48}$,
C.A.~Aidala$^{73}$,
Z.~Ajaltouni$^{5}$,
S.~Akar$^{59}$,
P.~Albicocco$^{18}$,
J.~Albrecht$^{10}$,
F.~Alessio$^{42}$,
M.~Alexander$^{53}$
% ...

{\footnotesize \it
% ...
$ ^{5}$Clermont Universit{\'e}, Universit{\'e} Blaise Pascal, CNRS/IN2P3, LPC, Clermont-Ferrand, France\\
% ...
$ ^{10}$Fakult{\"a}t Physik, Technische Universit{\"a}t Dortmund, Dortmund, Germany\\
% ...
$ ^{18}$INFN Laboratori Nazionali di Frascati, Frascati, Italy\\
% ...
$ ^{27}$Nikhef National Institute for Subatomic Physics, Amsterdam, Netherlands\\
% ...
$ ^{41}$Instituto Galego de F{\'\i}sica de Altas Enerx{\'\i}as (IGFAE), Universidade de Santiago de Compostela, Santiago de Compostela, Spain\\
$ ^{42}$European Organization for Nuclear Research (CERN), Geneva, Switzerland\\
% ...
$ ^{48}$H.H. Wills Physics Laboratory, University of Bristol, Bristol, United Kingdom\\
% ...
$ ^{53}$School of Physics and Astronomy, University of Glasgow, Glasgow, United Kingdom\\
% ...
$ ^{59}$University of Cincinnati, Cincinnati, OH, United States\\
% ...
$ ^{61}$Syracuse University, Syracuse, NY, United States\\
% ...
$ ^{73}$University of Michigan, Ann Arbor, United States, associated to $^{61}$\\
% ...
}
\end{flushleft}

\end{document}